\documentclass[a4paper,11pt]{article}
\pdfoutput=1

\usepackage{jcappub}
\usepackage{siunitx}

\usepackage[T1]{fontenc}
\usepackage[separate-uncertainty = true]{siunitx}

\title{Developing New Analysis Tools for Near Surface Radio-based Neutrino Detectors}

\author[a,m,1]{A.~Anker \note{Corresponding author}}
\author[b]{P. Baldi}
\author[a]{S. W. Barwick}
\author[c]{J. Beise}
\author[d]{D. Z. Besson}
\author[f]{P. Chen}
\author[a]{G. Gaswint}
\author[c]{C. Glaser}
\author[c]{A. Hallgren}
\author[h]{J. C. Hanson}
\author[i]{S. R. Klein}
\author[j]{S. A. Kleinfelder}
\author[e]{R. Lahmann}
\author[f]{J. Liu}
\author[f]{J. Nam}
\author[e,g]{A.~Nelles}
\author[a]{M. P. Paul}
\author[a]{C. Persichilli}
\author[e,g]{I. Plaisier}
\author[a]{R. Rice-Smith}
\author[k]{J. Tatar}
\author[e]{K.~Terveer}
\author[f]{S.-H Wang}
\author[a,l]{L. Zhao}

\affiliation[a]{Department of Physics and Astronomy, University of California, Irvine, CA 92697, USA.}
\affiliation[b]{Department of Information and Computer Science, University of California, Irvine, CA 92697, USA.}
\affiliation[c]{Uppsala University Department of Physics and Astronomy, Uppsala SE-752 37, Sweden.}
\affiliation[d]{Department of Physics and Astronomy, University of Kansas, Lawrence, KS 66045, USA.}
\affiliation[e]{ECAP, Friedrich-Alexander-Universität Erlangen-Nürnberg, 91058 Erlangen, Germany.}
\affiliation[f]{Department of Physics and Leung Center for Cosmology and Particle Astrophysics, National Taiwan University, Taipei 10617, Taiwan.}
\affiliation[g]{Deutsches Elektronen-Synchrotron DESY, 15738 Zeuthen, Germany. }
\affiliation[h]{Whittier College Department of Physics, Whittier, CA 90602, USA.}
\affiliation[i]{Lawrence Berkeley National Laboratory, Berkeley, CA 94720, USA.}
\affiliation[j]{Department of Electrical Engineering and Computer Science, University of California, Irvine, CA 92697, USA.}
\affiliation[k]{Research Cyberinfrastructure Center, University of California, Irvine, CA 92697, USA. }
\affiliation[l]{The William H. Miller III Department of Physics \& Astronomy, Johns Hopkins University, Baltimore, MD 21218, USA.}
\affiliation[m]{SLAC National Accelerator Laboratory, Menlo Park, CA 94025, USA}

\emailAdd{aanker@slac.stanford.edu, lzhao53@jhu.edu, sbarwick@uci.edu}

\abstract{ 
The ARIANNA experiment is an Askaryan radio detector designed to measure high-energy neutrino induced cascades within the Antarctic ice. Ultra-high-energy neutrinos above $10^{16}$ eV have an extremely low flux, so experimental data captured at trigger level need to be classified correctly to retain as much neutrino signal as possible. We first describe two new physics-based neutrino selection methods, or "cuts", (the updown and dipole cut) that extend the previously published analysis to a specialized ARIANNA station with 8 antenna channels, which is double the number used in the prior analysis. For a standard trigger with a threshold signal to noise ratio at 4.4, the new cuts produce a neutrino efficiency of > 95\% per station-year of operation, while rejecting 99.93\% of the background (corresponding to 53 remaining experimental background events). When the new cuts are combined with a previously developed cut using neutrino waveform templates, all background is removed at no change of efficiency.  In addition, the neutrino efficiency is extrapolated to 1,000 station-years of operation, obtaining 91\%. This work then introduces a new selection method (the deep learning cut) to augment the identification of neutrino events by using deep learning methods and compares the efficiency to the physics-based analysis. The deep learning cut gives 99\% signal efficiency per station-year of operation while rejecting 99.997\% of the background (corresponding to 2 remaining experimental background events), which are subsequently removed by the waveform template cut at no significant change in efficiency. The results of the deep learning cut were verified using measured cosmic rays which shows that the simulations do not introduce artifacts with respect to experimental data. The paper demonstrates that the background rejection and signal efficiency of near surface antennas meets the requirements of a large scale future array, as considered in baseline design of the radio component of IceCube-Gen2.  
}

\collaboration{ARIANNA collaboration}

\begin{document}
\maketitle
\flushbottom

\section{Introduction}
Neutrino astronomy in the ultra-high-energy regime (UHE, $E_\nu$>\SI{e17}{eV}) has the potential to provide insight into the sources of cosmic rays with energies in excess of \SI{e20}{eV} \cite{Fenu2017OO}. Neutrinos are ideal messengers because they have neutral charge and extremely low mass, so they travel through space without interacting with electromagnetic fields and have a low interaction probability. The challenge is measuring them since they rarely interact with matter, thus, a large target volume is needed to maximize the small neutrino flux. Many experiments have utilized ice as target material such as IceCube \cite{PhysRevD.98.062003} and radio-based pilot arrays \cite{barwick_glaser_2022} such as ARA \cite{PhysRevD.93.082003}, RNO-G \cite{RNOG}, and ARIANNA \cite{Gerhardt:2010js, BARWICK201512} (the focus of this paper). 

The ARIANNA experiment is comprised of two components: a hexagonal array of pilot stations with locations in Moore's Bay, Antarctica which is on the Ross Ice Shelf, and two additional stations located at the South Pole, Antarctica. The array at Moore's Bay was augmented by several special purpose stations that investigated a variety of improvements to the baseline technology of the initial hexagonal array. 
The ARIANNA hardware evolved into a versatile system based on the SST chip \cite{Kleinfelder2015,Barwick:2014rca, ARIANNA:2014fsk} that has been successfully used at sea level on the Ross Ice Shelf \cite{Anker:2019zcx}, and high elevation sites at the South Pole \cite{ARIANNA:2020zrg} and Mt. Melbourne in Antarctica \cite{TAROGE:2022soh}.  All but one of the deployed hardware systems operate autonomously, with independent power and communication.
The ARIANNA experiment has also given rise to the development of innovative simulation and reconstruction tools \cite{NuRadioReco, NuRadioMC,Glaser:2022lky} that were validated in in-situ measurements or through the measurement of cosmic rays that serve as a test beam for neutrino signals \cite{Barwick:2016mxm,Arianna:2021lnr,ARIANNA:2020zrg,GGaswintPhD, ARIANNAICRC2021Direction}.
A recent summary of previous results obtained by the ARIANNA experiment is provided in \cite{Glaser:2023akd}.

There are two signals of interest to these experiments: neutrino induced showers in ice and cosmic ray induced air showers. The detected neutrino signal (with frequency range \SI{50}{MHz} - \SI{1}{GHz}) is generated when UHE neutrinos collide with ice nuclei and create a chain reaction of particles cascades. These particle showers produce a time-varying charge excess in the shower front via the Askaryan effect \cite{Askaryan:1962hbi}. The detected cosmic ray signal is created in a similar way but the particle shower in air is elongated and thus dominated by geomagnetic emission \cite{Ardouin:2009zp}. The strength, frequency content, and radio pulse duration of cosmic rays are similar to the expected values for neutrinos \cite{Schoorlemmer_2016,Welling_2021}. Additionally, for the ARIANNA detector, the time dependent shape of electric fields generated by neutrinos and cosmic rays will have similar local effects on the antenna response. 

The ARIANNA pilot experiment was designed to search for UHE neutrinos with seven autonomous stations configured with just 4 downward facing log-periodic dipole antennas (LPDAs); however, several additional detector stations were constructed with 4 upward facing LPDAs to measure the flux of cosmic rays and further study detector capabilities, background rejection, and calibration techniques \cite{Anker:2019rzo,Beise:2022stx}. ARIANNA operated for about half a decade to explore and improve detector and analysis techniques. A few studies included improving the detector sensitivity of ARIANNA through a restricted bandwidth trigger \cite{Glaser_2021} and a deep learning filter \cite{Anker_2022}. Additionally, a search for neutrinos was done on 4.5 years of ARIANNA pilot station data and derived limits were found for UHE neutrino flux \cite{Anker:2019rzo}. That study used a template matching procedure to search for neutrino candidates in the data. The templates were produced by convolving the short duration (less than a few nanoseconds) bipolar electric field with the amplifier and LPDA antenna responses for a given arrival direction at the detector. The analysis in that earlier work exploited the fact that most background processes produce waveforms with quite different time dependence in the downward facing LPDA antennas (termed the "LPDA cut" in this paper). However, the expected sensitivity of future radio based high energy neutrino detectors are a factor of $10^4$ larger than demonstrated by the ARIANNA pilot array, which requires the development of additional analysis tools to meet the challenge of identifying neutrinos at high efficiency. 

This paper is focused on developing new analysis tools, or "cuts", that improve the neutrino efficiency, $\epsilon$, defined as the fraction of neutrino events observed by the detector that survive the analysis cuts, and applying these cuts to data collected by a specialized ARIANNA station with 8 antenna receivers, station 61, deployed at the South Pole in 2018. It has 4 more antenna receivers than the baseline ARIANNA station of reference \cite{Anker:2019rzo}.   This work describes three cuts that augment the analysis procedure of previously described analysis \cite{Anker:2019rzo} by incorporating the information from 4 additional antenna channels, which provide significant new information that simplifies the selection of neutrino events and rejection of background. We develop two physics-motivated cuts that, when combined with the previously developed LPDA cut, achieve $\epsilon$ > 90\% while residual background is reduced from  1 event per 20 station-years of operation to 1 event per 1,000 station-years. Then a new cut is developed that exploits deep learning tools. Deep learning networks offer the ability to train a neural network model on two distinct data sets to learn their distinguishing features. The success of deep learning in computer vision has led many physics experiments to adopt these analysis techniques \cite{baldi2021deep, Abbasi_2021,dl-baldi2014}. 

The ARIANNA detector stations used in this study will be described in more detail. The deep learning models and data sets used in this study will be defined. Then the neutrino analysis efficiency will be investigated for various cuts on the data, without the use of deep learning. Later neural networks will be used to determine the network output efficiency for experimental ARIANNA data. Lastly, a similar study is done with cosmic rays to provide a cross-check on the neutrino efficiency. The paper concludes with a summary of the results and a comment on future applications.

\section{ARIANNA and data sets}
For this study, a total of six data sets from the ARIANNA experiments are used. We first provide an overview of the experiment, before detailing each individual data set. As ARIANNA is a relatively small experiment, the data sets all have different properties and can be used for different purposes.

\subsection{The ARIANNA experiment}
\label{section:arianna_exp}
The ARIANNA experiment is primarily located on the Ross Ice Shelf in Moore's Bay, but also had two pilot stations located at the South Pole. Only a single station is needed to identify and reconstruct a potential neutrino signal, so stations can be deployed more flexibly in various configurations and locations. Although the baseline configuration of the ARIANNA station consists of 4 downward facing LPDA antennas, in this paper we analyze data from two specialized stations constructed with 8 antenna receivers: one located at the South Pole (station 61) and one located at Moore's Bay (station 52). Both stations produce similar data due to identical electronics and a similar hardware configuration consisting of 6 LPDAs, which are directional and primarily sensitive to horizontal polarization, and 2 fat-dipole antennas oriented vertically and therefore sensitive to vertical polarization.  An event was recorded when the waveform voltage for 2 of the 4 downward facing LPDAs exceeded $V_{th}$=4.4*$V_{\text{RMS}}^{\text{noise}}$ within a time interval of \SI{30}{ns}, where $V_{th}$ is the threshold voltage and $V_{\text{RMS}}^{\text{noise}}\sim$ \SI{10}{mV}, is the root mean square of the voltage fluctuations due to thermal noise. The $V_{\text{RMS}}^{\text{noise}}$ is determined in all antenna channels by a sample of events that were collected when the trigger was forced to occur at a random time.   A neutrino signal propagates within the ice medium at the speed of light, so the maximum time difference between any two LPDA antenna is \SI{30}{ns}, dictated by maximum geometrical separation of the downward facing LPDAs. Since the EM pulse from the neutrino interaction is approximately bipolar, the hardware trigger requires a positive and negative excursion within \SI{5}{ns} at the specified voltage threshold. The amplitude of the events is characterized by the signal to noise ratio, or SNR, and defined by SNR=$V_{max}$/$V_{\text{RMS}}^{\text{noise}}$, where $V_{max}$ is the largest absolute value of the signal amplitude of any of the 4 downward facing LPDA.

\begin{figure}[t]
\centering
  \includegraphics[scale=0.53]{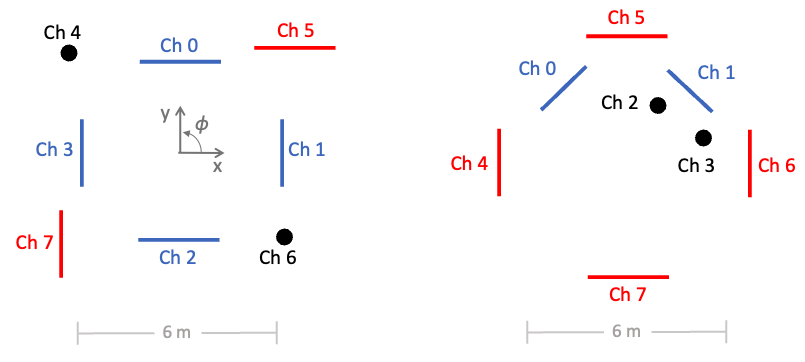}
  \caption{Overhead diagram of the station layouts for station 61 (left) and 52 (right) and their channel labels. Black dots are dipoles, blue lines are downward facing LPDAs, and red lines are upward facing LPDAs. The azimuthal angle of the arrival direction of the neutrino signal, $\phi$, is shown for station 61.}
  \label{fig:st_diagrams}
\end{figure}

 Station 61, located at the South Pole about \SI{5}{km} from the Amundsen-Scott Research Station, is configured to look for high energy neutrinos with a more advanced design than used by the baseline HRA detector station at Moore's Bay \cite{Anker:2019rzo}. The number of antenna channels was increased from 4 to 8, with 4 downward facing LPDAs, 2 upward facing with an orthogonal orientation of the LPDA planes, and 2 vertical fat dipole antennas (see the left diagram in \autoref{fig:st_diagrams}). This station was designed to contain all the antenna elements considered essential for the next generation surface detector. Station 52, located at Moore's Bay near the Antarctic coast, was configured to measure cosmic rays using 4 upward facing LPDAs and has a distinct layout as shown on the right side of \autoref{fig:st_diagrams}. Both stations rely on the same custom-built amplifiers, called series 300 amplifiers, and therefore have the same $V_{\text{RMS}}^{\text{noise}}$ around \SI{10}{mV}. For more information on the detailed architectures of the ARIANNA detector stations, refer to \cite{anker2020white,ARIANNA:2019scz}.

\subsection{Neutrino station 61}
The experimental data were collected by station 61 during the Austral summer months between December 10, 2018 and January 10,  2021, resulting in a total livetime of 1 year. Events were removed during periods of human activity at the detector site. Due to the diffuse neutrino flux limits set by IceCube and the duration of data taking from this one station, no neutrinos are expected in the data set. Thus, the data comprise primarily of thermal noise events, wind generated noise events \cite{triboelectric}, and noise generated from the ARIANNA electronics. In total, there are 74,530 (corresponding to a time averaged trigger rate of $\sim$ {$2\times 10^{-3}$} Hz)  experimentally triggered events in the data set from station 61, identified  by E-BG61. At the specified trigger threshold, events due to thermal fluctuations occur at a rate of $\sim 3\times 10^{-4}$ Hz, so the  majority of events collected by station 61 are non-thermal in nature.  

The simulated neutrino signal data set is generated with NuRadioMC \cite{NuRadioMC}, which simulates a representative set of expected neutrinos events for the ARIANNA detector. The randomly distributed events follow a GZK \cite{GZK_flux,GZK_2019} neutrino flux distribution with energies from $\si{10^{17}}{eV} - \si{10^{19.5}}{eV}$. There is also a weight cut of $10^{-5}$ performed so events with weights below this value are removed; since simulated neutrinos are generated uniformly in all directions, the weight cut removes lower probability neutrino events (corresponding to larger arrival angles below the horizon) that would not likely be measurable by ARIANNA. Next the radio pulses are propagated from the interaction vertex through the South Polar ice to the antennas at the detector station.  Then the ARIANNA station 61 detector is simulated with NuRadioReco \cite{NuRadioReco} with the exception that the dipole at channel 4 is buried at a depth of \SI{10}{m} instead of the actual depth of \SI{2.6}{m}. With this change, the generated neutrino template for the dipole of channel 4 is inappropriate for a neutrino search (and no results are presented here), but it does provide insight on the fraction of background events that would satisfy the dipole cut in a future station where the dipole will be deployed at \SI{10}{m} depth. The resulting neutrino radio signals are simulated in all 8 channels by convolving the electric field pulses with the antenna response. Though the signals in the shallow dipole of channel 6 are properly modeled, they are ignored in this analysis. Time delays introduced by the coaxial cables are subtracted in the data, so the simulated times do not include cable delays. Rayleigh noise within a bandwidth of 0 to \SI{1}{GHz} is superimposed on the neutrino pulses. After convolving with the amplifier response, a 10th order Butterworth bandpass filter between \SI{80}{}-\SI{500}{MHz} is applied to approximate the high pass response of the LPDA and low pass response of the readout electronics that digitizes the waveform at \SI{1}{gigasample/second}. The station is triggered if the signal pulse crosses a high and low threshold of 4.4 times the RMS noise, $V_{\text{RMS}}^{\text{noise}}$ within the 2 of 4 trigger logic criteria outlined in  \autoref{section:arianna_exp}. In total, 10,606 triggered events are in this set called station 61 simulated neutrinos, or S-NU61.

\subsection{Cosmic ray station 52}
The experimental data from station 52 are gathered from December 2018 to March 2019. Unlike the neutrino data set, these data contain tagged cosmic ray signal \cite{lzhao_2022,leshanICRC}, which are removed from this set. Therefore, this data set is expected to contain only background noise data. As with station 61, the station 52 experimental data contain thermal noise, wind generated noise, and noise generated from ARIANNA electronics. In total there are 97,955 events in this data set, which is called station 52 experimental data, or E-BG52. 

The extracted experimental cosmic ray data mentioned above is another data set in this analysis. Obtained from \cite{lzhao_2022,leshanICRC}, this data set was gathered by making various cuts on trigger rate, correlation, and arrival zenith. In total, there are 85 events in this set called experimental tagged cosmic rays, or E-CR52.

The simulated cosmic ray signal data set is created using the CoREAS software. CoREAS is a Monte Carlo code for simulating the radio emission from extensive air showers \cite{Huege_2013}. The CoREAS code contains cosmic rays over many different arrival directions and energies, but for this study, the events are re-weighted to match the expected energy and arrival direction distribution of the cosmic ray flux. Then, a detector simulation is performed using NuRadioReco \cite{NuRadioReco} where thermal noise is added to the cosmic ray signal and a \SI{80}{MHz} to \SI{500}{MHz} band-pass filter is applied. Only cosmic ray signal events that cross a high and low threshold of 4.4 times $V_{RMS}^{noise}$ within the 2 of 4 trigger logic criteria (see \autoref{section:arianna_exp}) are used. The simulation is most accurate for the 4 upward LPDAs, so the 2 downward facing LPDAs and dipole antennas are not used in this data set. Data set S-CR52 consists of 9,630 simulated cosmic rays.

The simulated thermal noise data set is similarly generated using NuRadioMC with the same trigger requirements and a threshold of 4.4 times the $V_{\text{RMS}}^{\text{noise}}$. In total, there are 50,000 triggered events in this set, called simulated thermal noise, or S-BG.

\begin{table}
\centering
\begin{tabular}{c  c  c} 
\hline \hline
 & Station 61 &  Station 52 \\
\hline
Simulated Signal & S-NU61: 10,606 & S-CR52: 9,630\\
Simulated Thermal Noise & - & S-BG: 50,000 \\
Experimental Data & E-BG61: 74,530 & E-BG52: 97,955 \\
Experimental Cosmic Rays & - & E-CR52: 85 \\
\hline\hline
\end{tabular}
\caption{Each data set is abbreviated when described in this analysis. The first letter denotes simulated or experimental data. The two letters after the dash are NU for neutrino, CR for cosmic ray, and BG for background noise data. The last two numbers are the station ID. Also given is the amount of events in each data set.}
\label{table:datasets}
\end{table}

\subsection{Contrasting the two ARIANNA stations used in this study}
\label{subsection:stn_comp}
Stations 61 and 52 have different configurations since they are searching for different particles. Station 52, with 4 upward facing and 2 downward facing LPDAs, was configured to search for cosmic rays by triggering on the upward facing LPDAs. Station 61 consists of 4 downward facing LPDA, 2 orthogonally-oriented upward facing LPDAs, and 2 vertically oriented fat dipoles. A summary of the configuration and corresponding channel numbers for each station are given in \autoref{table:chan_assignment}. Data was selected for the periods of time when the trigger was configured to search for neutrinos by requiring signals to be present in at least 2 of the 4 downward facing LPDAs. 

\begin{table}
\centering
\begin{tabular}{c  c  c} 
\hline \hline
 & Station 61 &  Station 52 \\
\hline
ch0 & downward LPDA & downward LPDA\\
ch1 & downward LPDA & downward LPDA\\
ch2 & downward LPDA & dipole, 4.5 m depth\\
ch3 & downward LPDA & dipole, 8.5 m depth\\
ch4 & dipole, 2.6 m depth & upward LPDA\\
ch5 & upward LPDA & upward LPDA\\
ch6 & dipole, 2.6 m depth & upward LPDA\\
ch7 & upward LPDA & upward LPDA\\
\hline\hline
\end{tabular}
\caption{The antenna configuration and corresponding channel numbers for station 61 and 52. Downward LPDA refers to the LPDAs that are downward facing and vice versa for upward LPDA. }
\label{table:chan_assignment}
\end{table}

\begin{figure}[t]
\centering
  \includegraphics[scale=0.27]{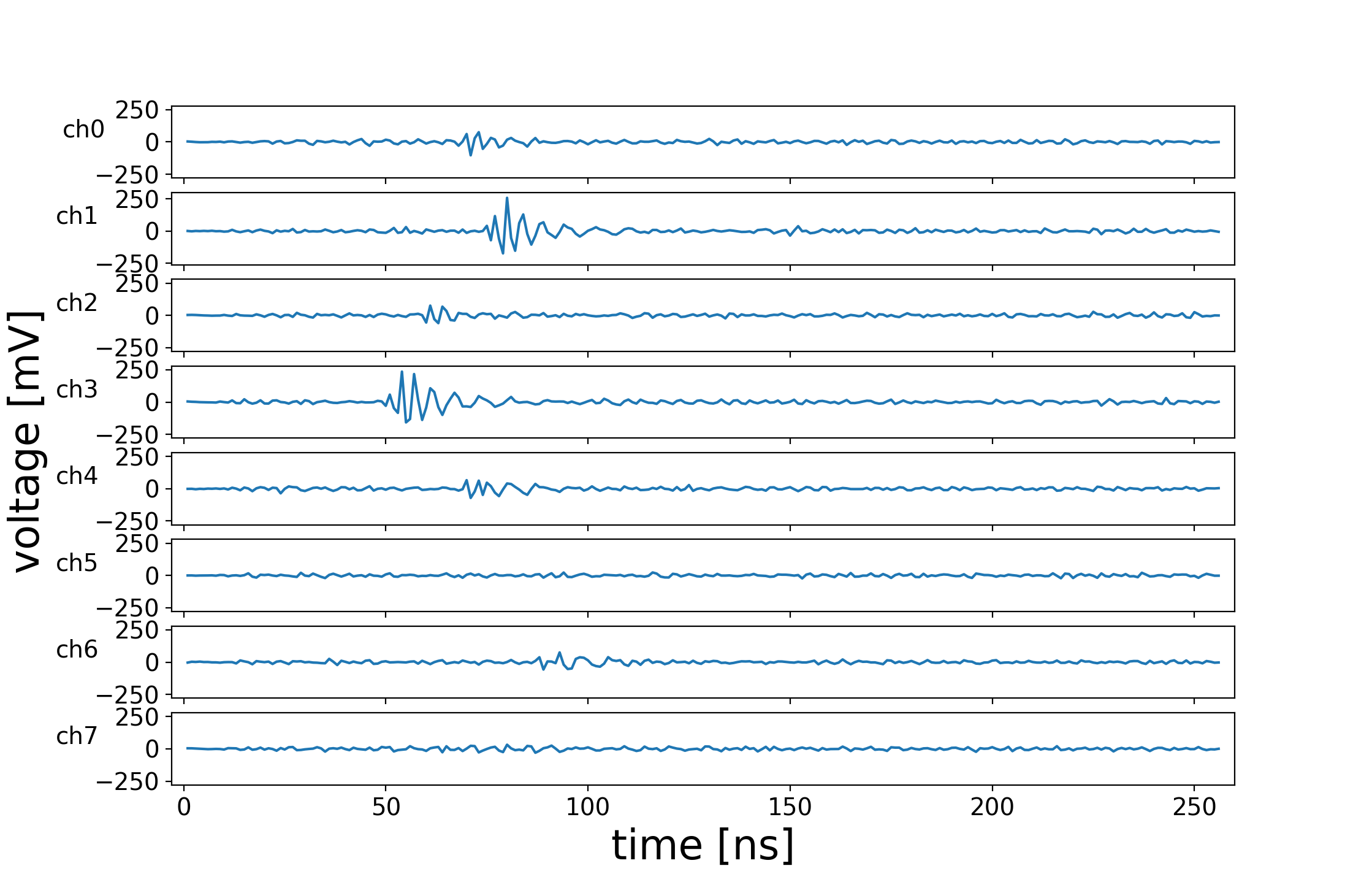}
  \includegraphics[scale=0.27]{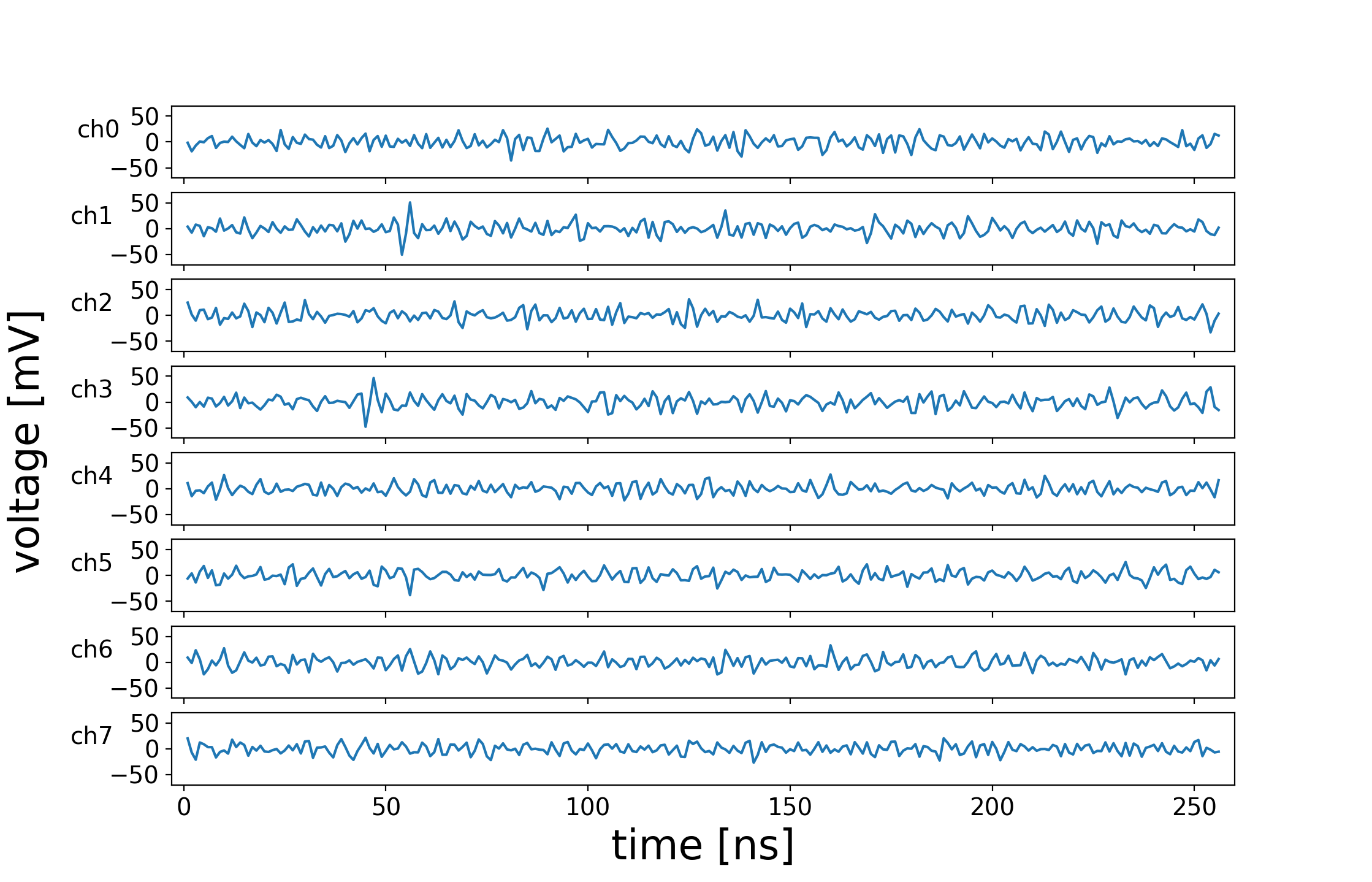}
  \includegraphics[scale=0.27]{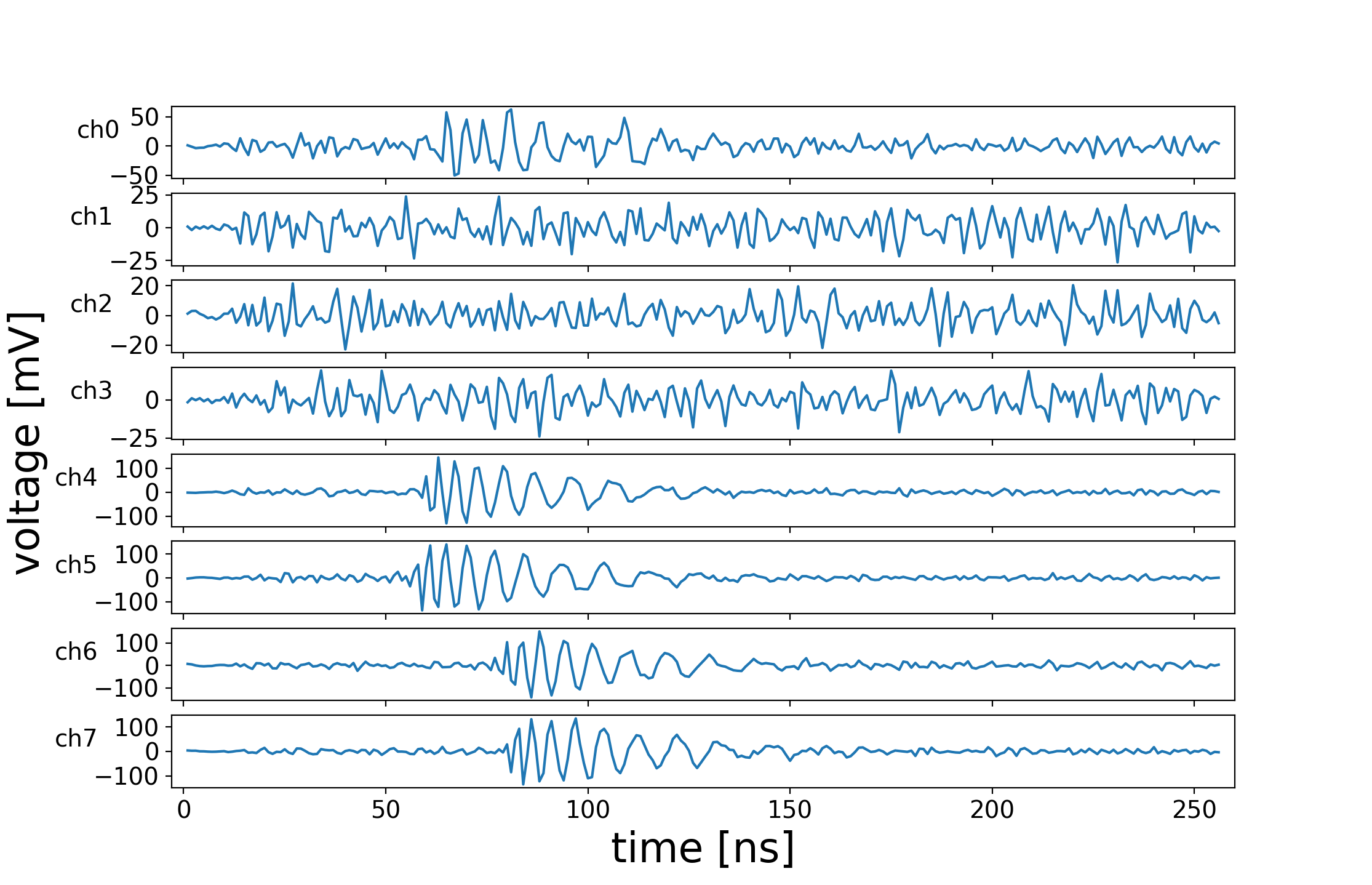}
  \includegraphics[scale=0.27]{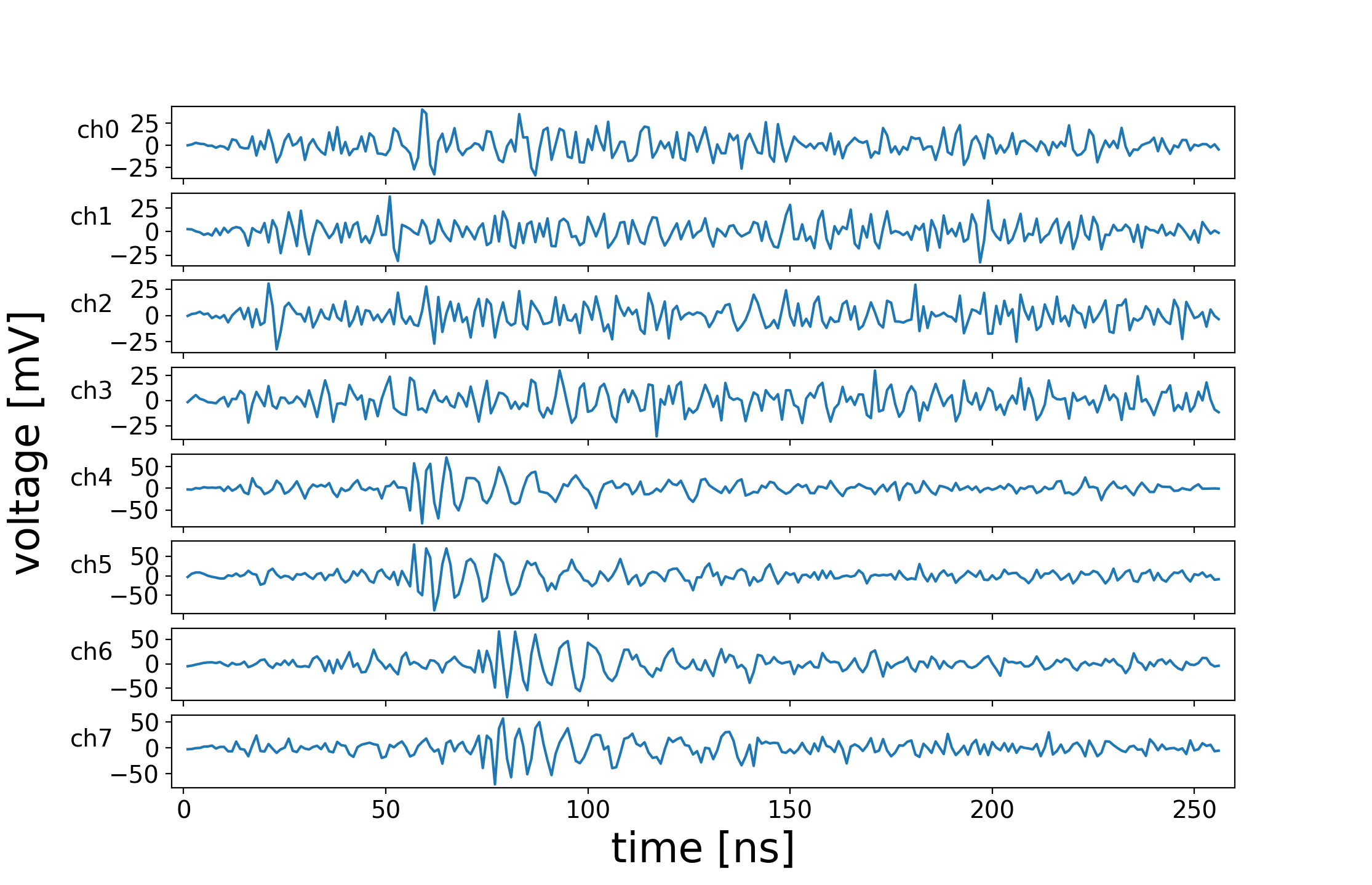}
  \caption{Waveforms for a representative simulated neutrino (top left, station 61), thermal noise event (top right, station 61), simulated cosmic ray (bottom left, station 52), and experimental cosmic ray (bottom right, station 52) with approximately the same arrival direction as the simulated cosmic ray. Note that station 52 is triggered on the last four channels whereas the two station 61 waveforms are triggered on the first four channels. For a complete list of channel assignments, see \autoref{table:chan_assignment}.}
  \label{fig:all_wfs}
\end{figure}

 There are two reasons to expect that cosmic ray and neutrino waveforms will show similar structure. First the time dependent electric fields generated by cosmic ray and neutrino interactions are both of very short duration (around a few nanoseconds) compared to the response of the antennas and amplifiers. Second both stations use identical amplifier systems. The similarity in shape is rather striking compared to thermal noise fluctuations that commonly trigger the station; see \autoref{fig:all_wfs}. There are few notable differences between the waveforms observed by station 52 and those measured by station 61. First, station 52 observes cosmic rays with upward facing LPDAs that point in the same direction as anthropogenic and weather-induced backgrounds that propagate through the atmosphere. In contrast, the backgrounds in the data collected by station 61 arrive in the direction of the back lobe of the downward facing LPDA antenna, a significantly less sensitive direction with different frequency characteristics compared to the front lobe. Consequently, the waveforms of background events from high winds and anthropogenic sources will more closely match the typical waveforms produced by cosmic ray events.

Geographically, there are additional differences between the two stations. Station 61 is located at the South Pole at an elevation of \SI{2800}{m}, about \SI{5}{km} from South Pole Station. Station 52 is located at sea level at Moore's Bay on the Ross Ice Shelf, Antarctica, more than \SI{100}{km} from the nearest research base. Due to its proximity to the South Pole, station 61 observes more anthropogenic noise from weather balloon launches, spark plug noise from snowmobiles and other vehicles with motors, and communications to aircraft and other bases in Antarctica. Station 52 is more remote, so these sources of noise are comparatively small. However, the average winds are stronger at Moore's Bay, so wind generated radio noise from blowing snow particles is more prevalent. In addition, the electrical noise from the battery charging electronics on the ARIANNA station is greatest during the time of year when the position of the sun transitions from above the horizon to below the horizon. At Moore's Bay, this happens daily during sunrise and sunset for a period of about a month whereas at the South Pole, sunrise and sunset occurs only once per year. Therefore, the experimental data from station 52 will contain a relatively high fraction of large amplitude background events from electronic noise compared to station 61. The ARIANNA neutrino detection efficiency will be investigated first in \autoref{sec:eff_cuts} with variable-based cuts and then later in \autoref{sec:nu_search} with deep learning.

\section{Evaluating the neutrino analysis efficiency without Deep Learning}
\label{sec:eff_cuts}

The ARIANNA collaboration published an upper limit on the flux of neutrinos with energy, $E_{\nu}>10^{17}eV$ that relied on data from ARIANNA pilot stations with just 4 antenna channels, all devoted to downward facing LPDA antennas \cite{Anker:2019rzo}. The neutrino efficiency (defined in that paper as the fraction of neutrino events in the raw data that remain after applying the cuts to reduce the expected background to 0.5 events in the full data extending over 4.5 years) was greater than 0.8. That analysis relied on first selecting time dependent waveforms from the pair of parallel downward facing LPDAs with the largest amplitude and then cross-correlating them with the expected shape from neutrino events, or "templates". They were obtained from widely-available simulation tools such as NuRadioMC \cite{NuRadioMC}. Unsurprisingly, the correlations were stronger for larger amplitude events, so the analysis procedure defined a signal region in the cross-correlation versus signal amplitude plane. Although this work was encouraging, the integrated livetime from that pilot array was less than 1\% of the requirement for future generation arrays such as IceCube-Gen2\cite{Gen2Radio2021}. New analysis tools are required to increase the analysis efficiency for high energy neutrinos when the integrated background increases by a factor of 100 or more. 

\begin{figure}[t]
\centering
  \includegraphics[scale=0.4]{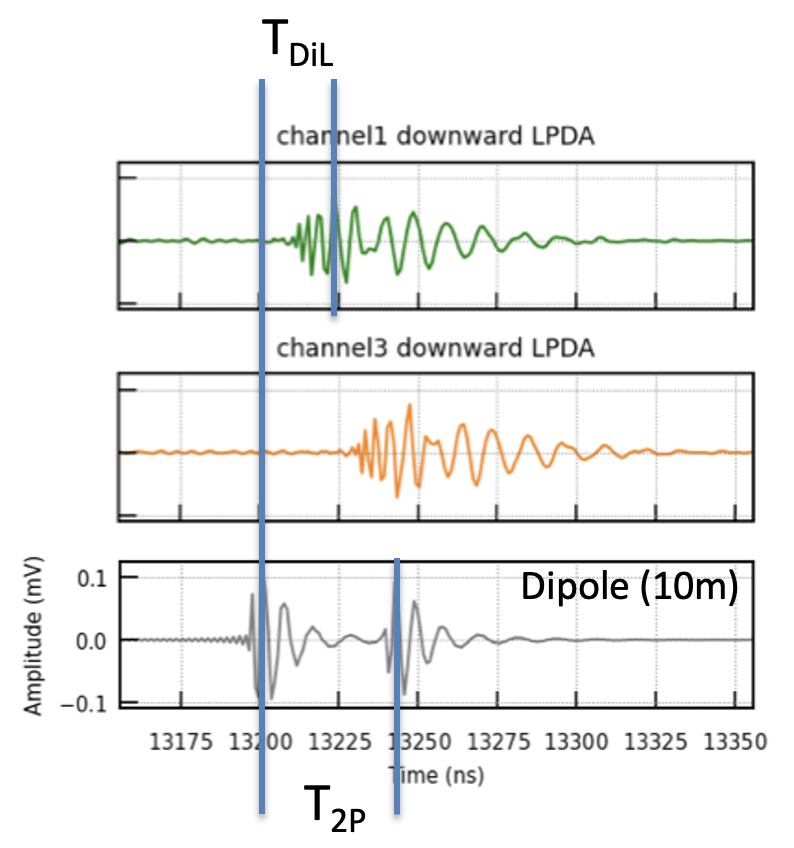}
  \includegraphics[scale=0.29]{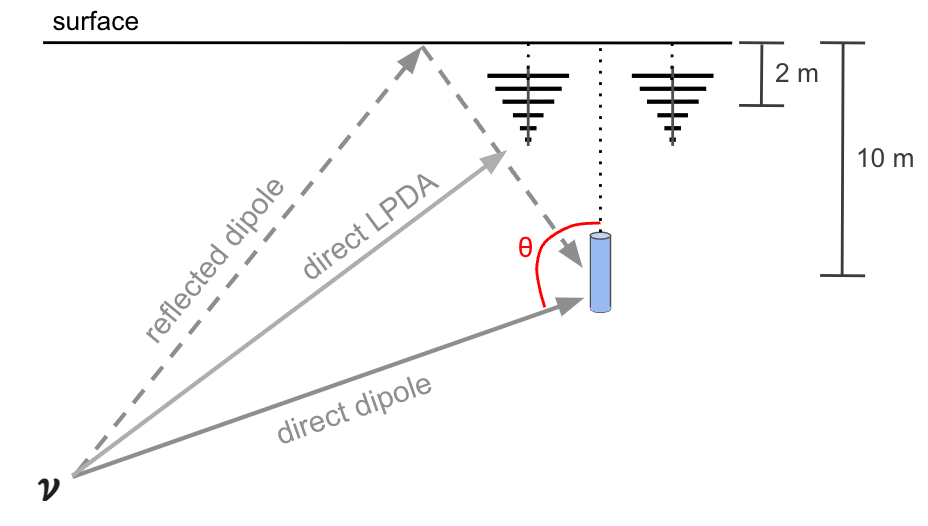}
  \caption{(left) Simulated neutrino waveforms observed by a vertically oriented fat dipole buried \SI{10}{m} beneath the snow surface (bottom waveform), and two parallel downward facing LPDAs (top and middle waveform). The relative timing between the fat dipole and earliest arriving LPDA signal, $T_{DiL}$ is shown, as is the relative timing between the direct and reflected pulse in the dipole channel, $T_{2P}$. (right) A sketch of the propagation paths of radio signals from the particle cascade induced by a neutrino interaction ($\nu$ symbol) to the dipole and closest LPDA antenna of ARIANNA station 61. Note that the sketch is not drawn to scale. For the vast majority of neutrino events, the direct LPDA ray, the direct dipole ray, and (prior to striking the surface) the reflected dipole ray are nearly parallel to each other.}
  \label{fig:DipoleWaveform}
\end{figure}

ARIANNA station 61 was augmented to read 8 antenna channels. In addition to the 4 downward LPDAs, the station includes: (1) two upward facing LPDAs to identify cosmic rays, wind, and anthropogenic backgrounds that propagate through the air, and (2) two fat dipoles to observe signals with vertical polarization. It is expected that a future station will include at least one dipole buried to a depth of $\sim$\SI{10}{m} to observe the distinct neutrino signature of a double pulse from a neutrino event (the second pulse is due to reflections off the firn-air surface and is only produced by radio emission from within the ice) \cite{Anker:2019zcx}. The right panel of \autoref{fig:DipoleWaveform} provides a sketch of the relevant signal propagation paths from the shower cascade of the neutrino interaction. The firn-air surface is assume flat so that the reflection angle is the same as the incidence angle. Since the typically large distance to the neutrino vertex (> 200 m) is much larger than the distances between the antenna elements on a single detector station, the direct dipole and LPDA rays, and the reflected dipole ray prior to reaching the surface, are nearly parallel to each other. Due to this feature, the reconstructed arrival direction of the direct LPDA ray (the arrival zenith angle is indicated by $\theta$)  can be used as a proxy for the arrival direction of the direct dipole ray. The left panel of \autoref{fig:DipoleWaveform} shows an example of the relative timing of the waveforms generated by a station with downward facing LPDAs and a fat dipole buried at a depth of \SI{10}{m}. The time difference between the two pulses in a dipole channel is $T_{2P}$, the signal time difference between the earliest arrival time in the LPDA and the first pulse in the dipole is $T_{DiL}$. Anthropogenic and cosmic ray radio signals that propagate through the atmosphere prior to entering the ice do not generate double pulse structures with time delays smaller than \SI{100}{ns}.  

The double pulse structure from direct and reflected signal trajectories was cleanly observed in a dipole buried at 8.5m in station 52 \cite{Anker:2019zcx}. In that work the double pulse structure was used to measure snow accumulation on the surface. The fat dipoles in station 61 are not buried as deep (about \SI{2.6}{m}), so a distinct double pulse structure from a neutrino event is not observable.  However, these data provide crucial information on the rate of background events that mimic the double pulse signature.  

A conventional way to identify neutrino events is to develop selection criteria or "cuts" that focus on event characteristics that are significantly different between neutrino and background noise events. In addition to the "cross-correlation cut" that was developed in the previously reported analysis \cite{Anker:2019rzo}, two new cuts are developed that rely on new information from the upward facing LPDAs and dipole antennas in station 61.  In this section,  we describe the new updown and dipole cuts and report on the overall neutrino efficiency of the combined cuts, defined as the percentage of simulated neutrino events passing a particular cut or series of cuts relative to the number of neutrinos in the data prior to the application of the cuts. The particular cut thresholds are adjusted to reduce the expected number of background events to <0.1 events in the experimental data. Future work will optimize the cut thresholds.

\subsection{Updown cut}
Neutrino interactions and the subsequent generation of radio signals originate from below the firn-air interface in most cases. On the other hand, many types of backgrounds, including wind-induced and anthropogenic events, arrive from above the ice surface. Due to the directionality of the LPDA antennas, events from above the surface typically have a larger signal in the upward facing antennas than downward facing ones, and vice versa. By studying the ratio between the amplitudes in upward and downward facing antennas, a cut was developed to reject above-the-surface events while efficiently retaining below-the-surface events.

NuRadioMC \cite{NuRadioMC} was used to simulate neutrino signals in the ARIANNA stations. The energy spectrum was obtained from a widely discussed cosmogenic neutrino model \cite{GZK_2019}. The trigger consisted of first applying a channel specific requirement that the roughly bipolar waveform amplitude exceed $4.4\times V_{rms}$ within \SI{5}{ns}, where $V_{rms}$ is the root mean square of the random voltage fluctuations induced by thermal noise. Then at least 2 of the 4 downward facing LPDAs must satisfy the channel specific requirement within \SI{30}{ns}. These two trigger requirements replicated the experimental conditions. A bandpass filter (\SI{80}{MHz} - \SI{500}{MHz}) was applied to both the experimental waveforms and the simulated waveforms after the amplifier response was included. The ice model was based on the site of the ARIANNA station. For station 61 (52), the South Pole (Moore's Bay) ice model was used in the simulation. 

\begin{figure}[t]
\centering
  \includegraphics[scale=0.28]{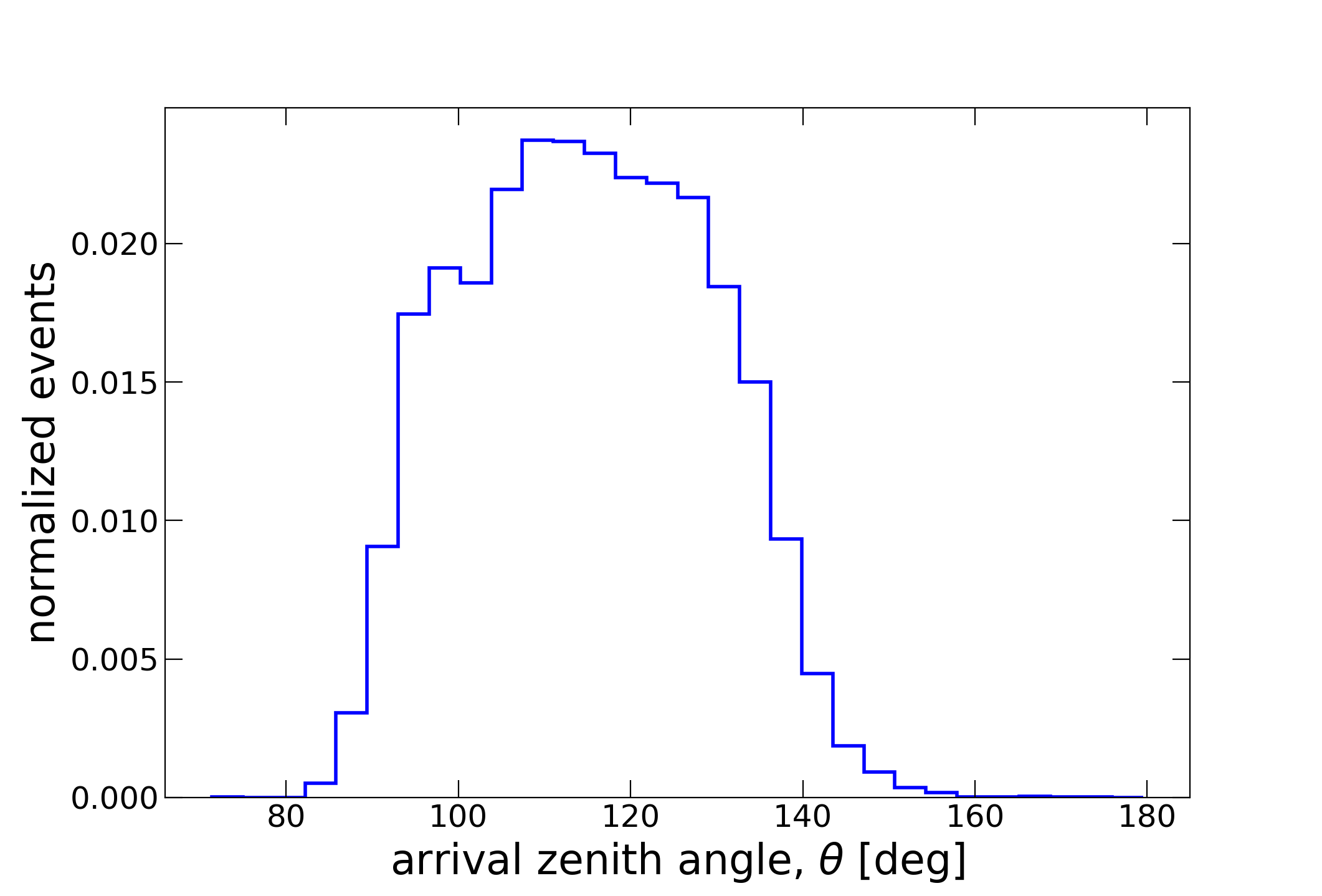}
  \includegraphics[scale=0.29]{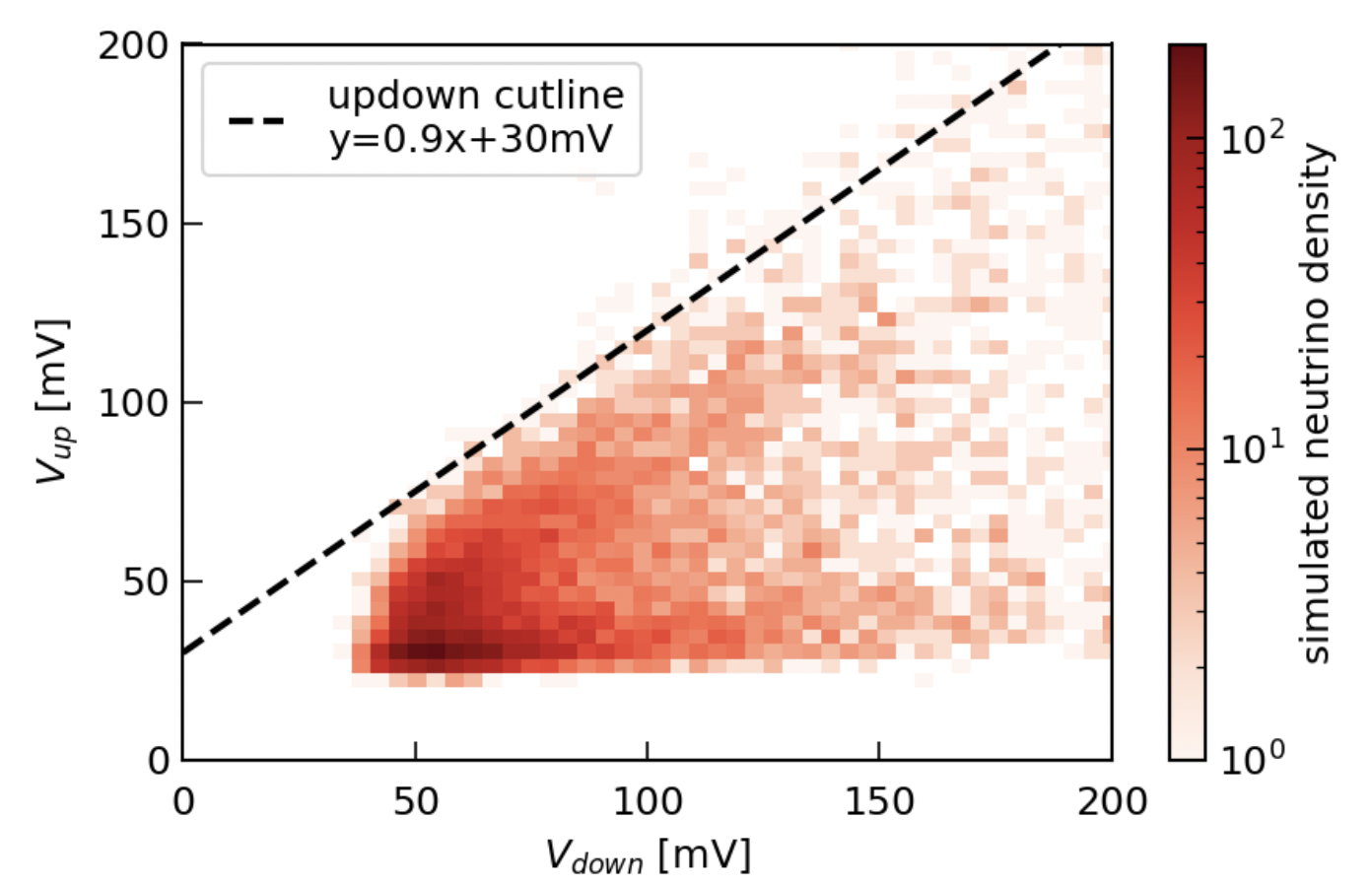}
  \caption{(left) Normalized distribution for the zenith angle of arrival direction, $\theta$ from neutrino events in South Pole ice. The coordinate system defines \SI{180}{deg} for a propagating direction that is straight up. (right) The maximum voltage of the downward LPDAs ($V_{down}$) versus upward LPDAs ($V_{up}$) for the weighted simulated neutrino density. The dashed black line defines the updown cut, which retains 99\% of the simulated neutrino events.}
  \label{fig:updwn}
\end{figure}

The directional (or "updown") cut exploits the intrinsic directional capabilities of the LPDAs. It is relatively common for sources of pulsed radio background to have larger signals in the upward facing LPDA, whereas for neutrino signals, the pattern is reversed.  However, the difference in signal strength between upward and downward facing LPDA is mitigated for neutrino signals arriving near the horizon because the antenna response is very similar at small angular differences, both positive and negative. Another mitigating effect concerns reflections at the firn-air interface. The left panel of \autoref{fig:updwn} shows that the mean arrival angle of neutrino radio signals is \SI{110}{deg} which is well within the region of total internal reflection that begins at \SI{130}{deg}. A little more than 87\% of the reflected rays from neutrino events will undergo total internal reflection (TIR). For neutrino signals arriving at angles that generate significant reflected power, upward facing LPDAs will observe radio signal. The "updown" cut is indicated by the dashed line in the right panel of \autoref{fig:updwn}, where $V_{down}$ is defined as the maximum positive voltage of any of the downward facing LPDAs, and  $V_{up}$ is defined as the maximum positive voltage of either of the two upward facing LPDAs. Events below the cutline in the $V_{down}$-$V_{up}$ plane are retained for further analysis. This step of the analysis procedure reduces the background event sample from 74,530 events to  41,821 events, while $\epsilon_{updown}$ >99\% for the simulated neutrino signal. 

\subsection{Dipole cut}
Due to the expected double pulse structure for neutrino events, a fat dipole buried at \SI{10}{m} offers important new information that provides a powerful tool to improve the background rejection while selecting neutrino events at high efficiency. The radio signal from the neutrino interaction that directly propagates to the antenna creates the first pulse of the dipole waveform. The time delayed second pulse is due to the reflection from the firn-air interface at the surface. As discussed later, this feature can be observed for most neutrino events because radio pulses from neutrino events typically reach at the snow surface with arrival angles (where \SI{180} degrees is vertically up-going) between \SI{90} deg and \SI{130} deg, the largest angle corresponding to total internal reflection (TIR), and the amplitudes of the direct and reflected pulses are expected to be similar due to the symmetry of the response of the dipole antenna relative to the normal direction from the dipole axis. For arrival angles between \SI{130}{deg} - \SI{143}{deg} , reflection is not total, and the relationship between the amplitude of the direct pulse and second pulse is calculated from the Fresnel equations. Approximately 11.4\% of the neutrino events fall into this category. The ratio of the amplitudes of the direct and delayed pulse depends on arrival angle, which is well measured by the surface station \cite{ARIANNA:2020zrg}, and polarization, which is not included in this analysis. Finally, for arrival angles greater than \SI{143}{deg}, a delayed pulse is not included in the template because the reflected amplitude is small. Fewer than 1.5\% of the neutrino events arrive at these angles.  

 The dipole cut procedure requires a neutrino template waveform for the fat dipole antenna to cross-correlate with the observed waveform. The template encodes the expected time delay ($T_{2P}$) and fractional amplitude of the delayed pulse as a function of arrival angle.  Finally, the relative time between the direct pulse in the dipole and earliest arriving signal in the LPDA channels ($T_{DiL}$) is computed and included in the template analysis by restricting the time window of cross-correlation calculation.  The next section describes the procedure to generate the dipole template.  

\subsubsection{Generating the dipole template}
The primary task in generating a dipole template involves estimating $T_{2P}$, $T_{DiL}$ and the relative amplitude between the first and second pulse. These three quantities depend on the signal arrival direction, which differs from event to event. As a result, a unique template has to be generated for each event to account for the difference in arrival directions.

\begin{figure}[t]
\centering
  \includegraphics[scale=0.28]{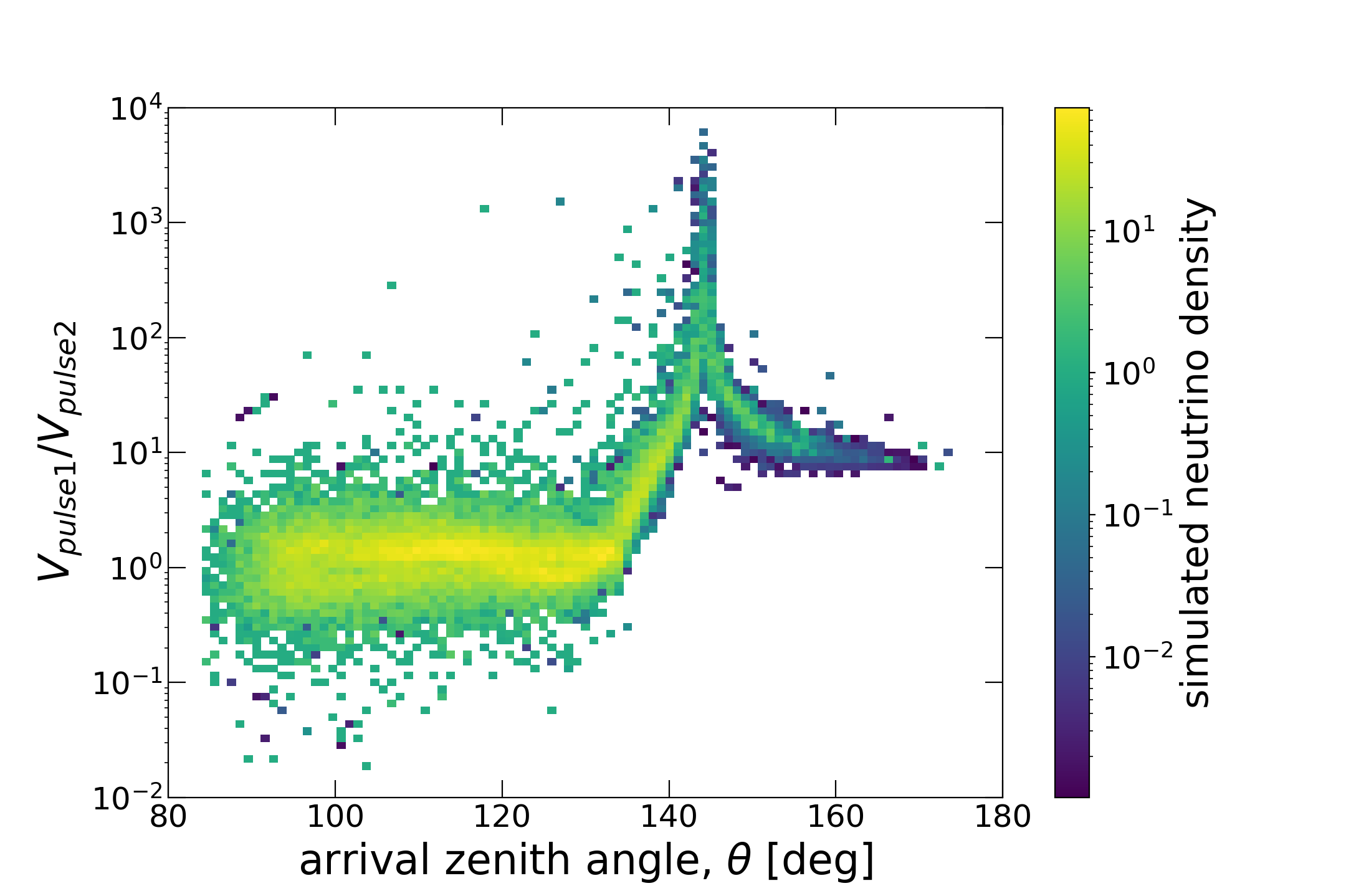}
  \includegraphics[scale=0.28]{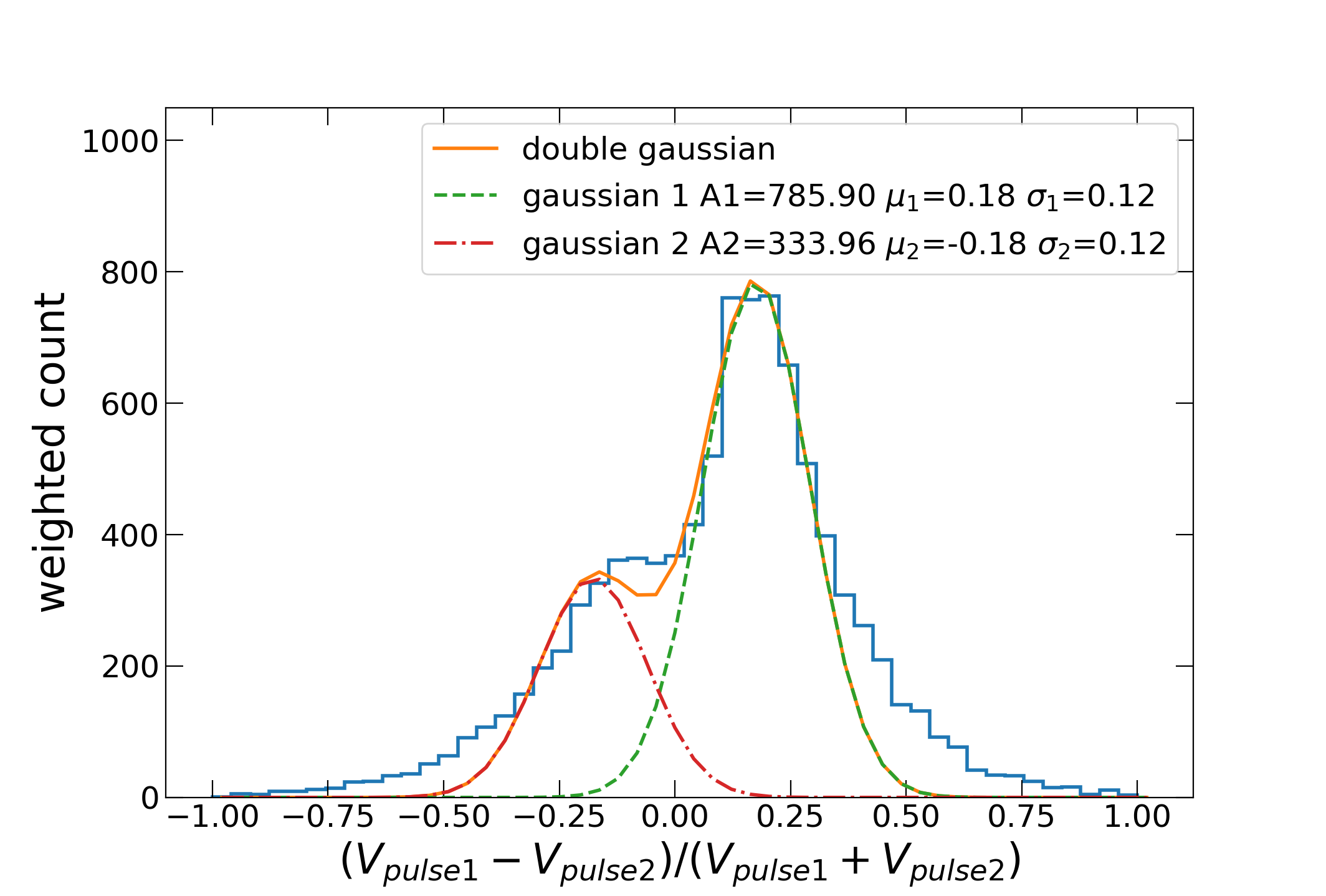}
  \caption{(left) Density plot of the ratio of pulse amplitudes in the dipole for simulated neutrino signals as a function of the zenith angle of the arrival direction at the ARIANNA station, where $V_{pulse1}$ is the amplitude of the earliest pulse in the dipole waveform and $V_{pulse2}$ is the amplitude of the delayed pulse.  The feature at \SI{143}{degrees} is due to events arriving at the Brewster angle. (right) Histogram of neutrino counts as a function of the amplitude asymmetry for zenith angles less than \SI{130}{degrees} (which corresponds to TIR). The majority of events show a positive asymmetry fraction of 0.18 $\pm$ 0.12 (green). Also shown is the summed double Gaussian pulse (orange).}
  \label{fig:dcut}
\end{figure}

Although the characteristics of the neutrino pulse shape from the dipole depend on physical factors such as the emission angle relative to the Cherenkov cone, the template construction is simplified by selecting a time dependent waveform from a typical emission angle of \SI{2}{degrees}. The delayed pulse due to reflection (pulse 2) is identical in shape but with the amplitude re-scaled according to its arrival zenith angle. The re-scaling done on pulse 2 accounts for the change in amplitude after reflection from the snow surface. This approximation relies on the assumptions that the signal is a plane wave, and the snow surface behaves as a perfect mirror. With these assumptions, there is negligible variation in the emitted electric fields between the direct and reflected pulses due to slightly different emission angles relative to the Cherenkov cone. The left panel of \autoref{fig:dcut} shows the behavior of the amplitude ratio of the earliest pulse ($V_{pulse1}$) to the amplitude of the delayed pulse ($V_{pulse2}$). The amplitude ratio was fit by a piece-wise function over three zenith angle intervals, and summarized in \autoref{tab:DipoleAmpRatio}.

\begin{table}
  \begin{center}

    \begin{tabular}{ll} 
      \hline
      \hline
      \textbf{Zenith Interval (deg)} & \textbf{Amplitude Ratio} \\
\hline
      $\theta$ < 130 & 1 \\
      130 < $\theta$ < 143 & $(0.19e^{0.64(\theta -133.5)}+1)^{-1}$ \\
     $\theta$ > 143 &  0 \\
           \hline
      \hline
    \end{tabular}
  \end{center}
      \caption{Amplitude ratio ($V_{pulse2}/V_{pulse1}$) as a function of zenith angle of the arrival direction}
    \label{tab:DipoleAmpRatio}
\end{table}

For zenith angles smaller than \SI{130}{degrees}, the TIR approximation of equal amplitudes is violated by 18\%, as shown in the right panel of \autoref{fig:dcut}. This asymmetry is due to the fact that for most interaction vertices, the emission angle relative to the Cherenkov cone is both outside of the Cherenkov cone and slightly smaller for the direct ray than the reflected ray. Since the emission amplitude decreases with increasing emission angle, the direct pulse will be larger than the reflected.  However, about 29\% of the neutrino events show a negative asymmetry because the direct and reflected rays are emitted from the interior of the Cherenkov cone. Consequently, the amplitude of the electric field of the reflected ray could be larger at the emission vertex than the direct ray. The distribution of cross-correlation values the dipole antenna, $\chi_{Di}$, is not very sensitive to the exact amplitude ratio, with nearly equivalent results for ($V_{pulse1}$)/($V_{pulse2}$) = 0.8 - 1.2. Future work will investigate the inclusion of the measured emission angle and vertex location to produce more accurate predictions of the amplitude ratio in the dipole template.

\begin{figure}[t]
\centering
  \includegraphics[scale=0.45]{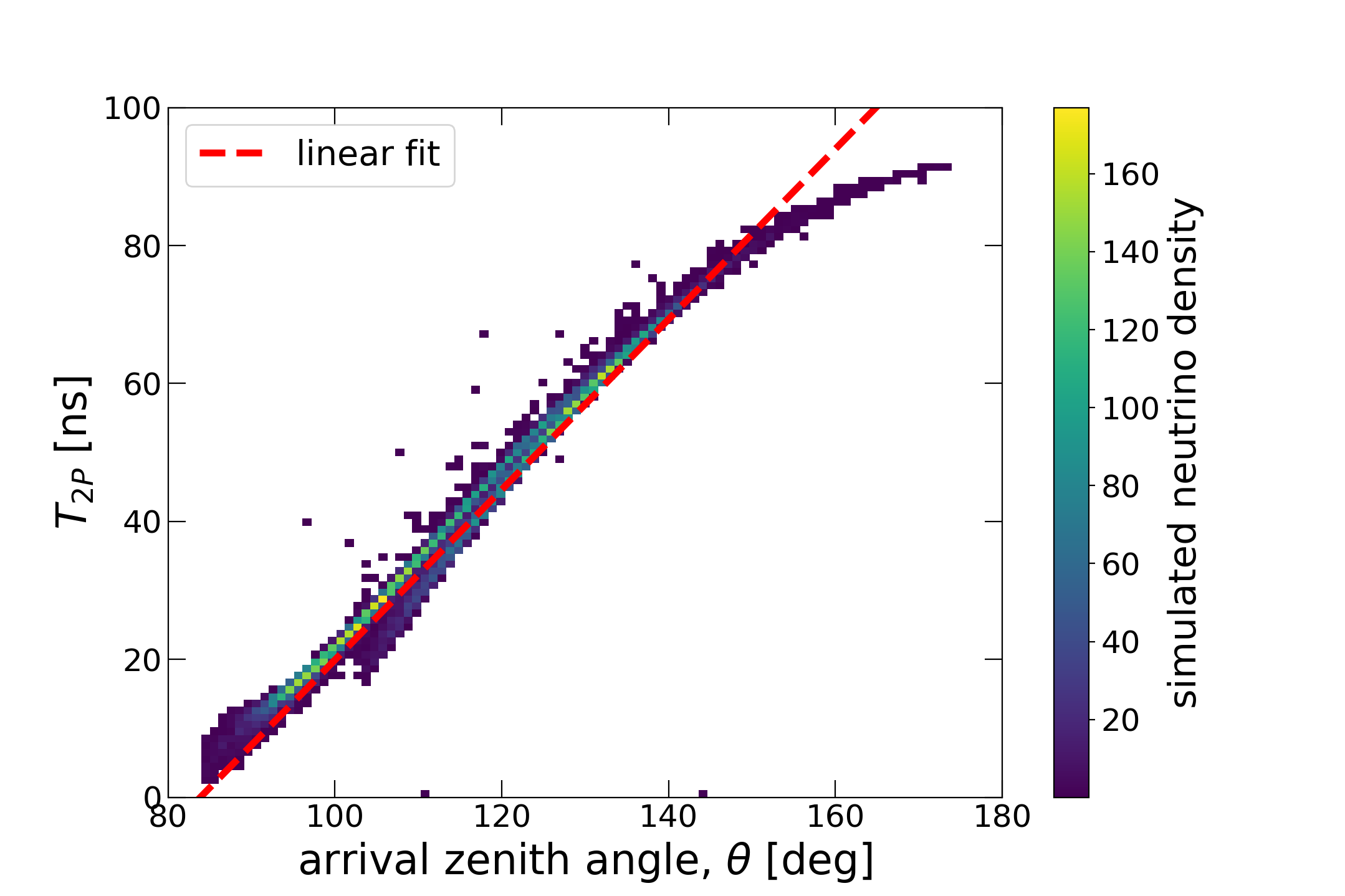}
  \caption{Density plot of the time interval, $T_{2P}$, between neutrino-induced pulses in the dipole antenna, buried at a depth of \SI{10}{m}, as a function of the zenith angle of the arrival direction at the LPDA.}
  \label{fig:dt_z}
\end{figure}

The time difference between the direct and reflected pulse in the dipole waveform depends on the cosine of the arrival direction angle, $\theta$. The largest time difference occurs for signals propagating straight up through the ice ($\theta = 180$ deg). Since most radio signals from astrophysical neutrinos arrive with zenith angles between 90 and 130 deg, a fit using a cosine function was unnecessary. To simplify the calculation of time delay between the two pulses in the dipole template, a simple linear fit to the simulated neutrino distribution shown in \autoref{fig:dt_z} gives  $T_{2P} =(1.234\theta -103.4)$ ns, where $\theta$ is the zenith angle of the arrival direction at the ARIANNA station, in degrees. The red line in  \autoref{fig:dt_z} shows the fit and range of validity, which includes most of the neutrino signals.     Pulse 2, after amplitude scaling,  is delayed by $T_{2P}$ and superimposed on pulse 1. For arrival directions where $T_{2P}$ is small, pulse 1 and 2 interfere. To accommodate the uncertainty in the estimated $T_{2P}$, a set of templates are created that adjust $T_{2P}$ by values ranging from \SI{-10}{ns} to +\SI{10}{ns}. To improve accuracy and account for the small errors in the fit, the raw waveforms are upsampled from \SI{1}{sample/ns} to \SI{10}{samples/ns}, and the time delay to Pulse 2 is incremented by \SI{0.1}{ns} steps on the upsampled waveform. 

The left panel of \autoref{fig:dipcorr} compares a representative example of the dipole template with the simulated waveform for a neutrino event with an arrival direction of \SI{138}{deg} (which is in the angular region where the reflection off the surface transitions from small to total). Both waveforms are noiseless. The cross-correlation value of 0.94 for this event indicates that the predicted template provides a good but not perfect approximation of the simulated signal. For this event in particular, the assumption that the second pulse is identical to the first in shape does not hold due to the phase shift after reflection and the emission at a slightly different angle with respect to the Cherenkov cone.  

\subsubsection{Cross-correlation using dipole template}
\begin{figure}
\centering
  \includegraphics[scale=0.27]{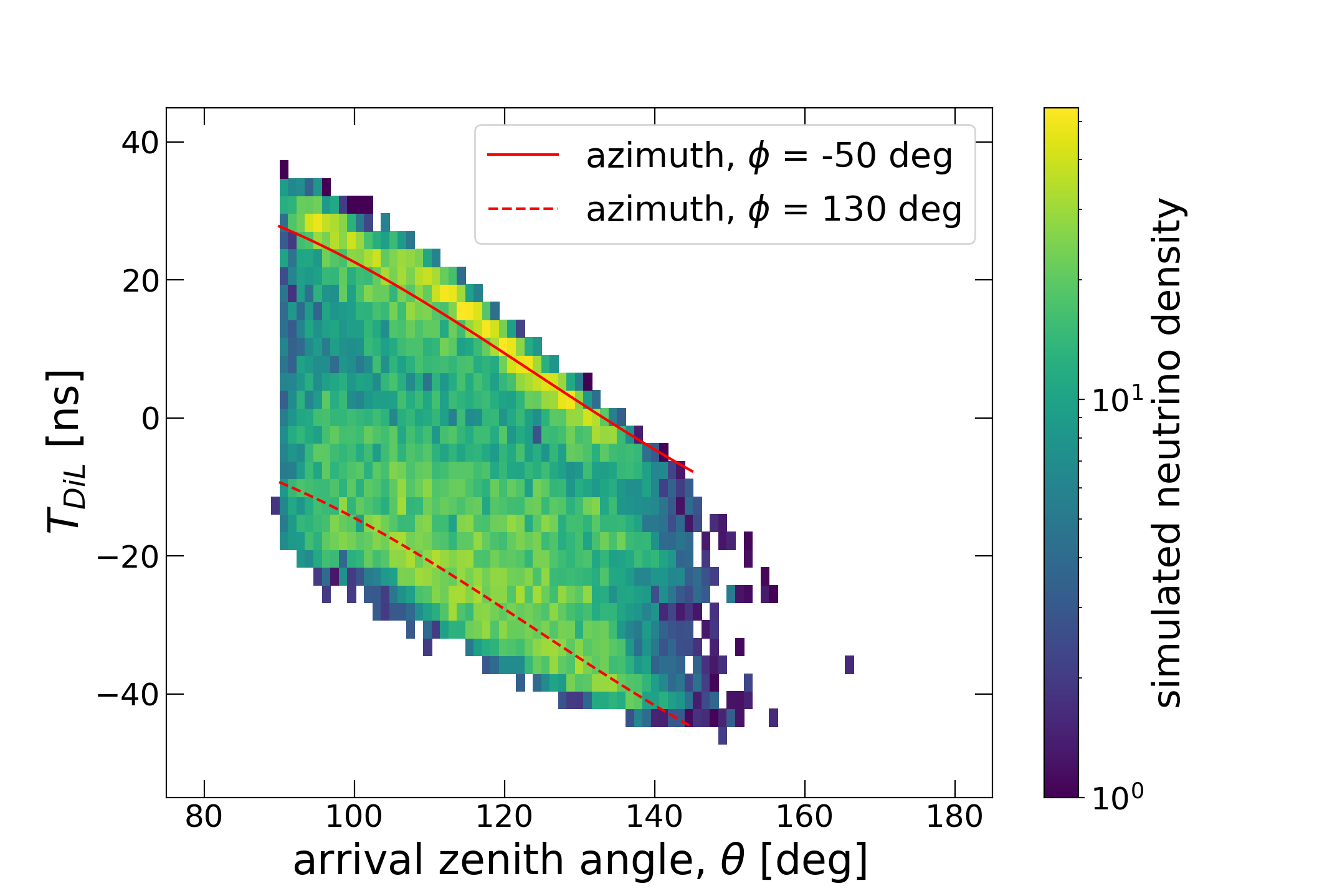}
   \includegraphics[scale=0.27]{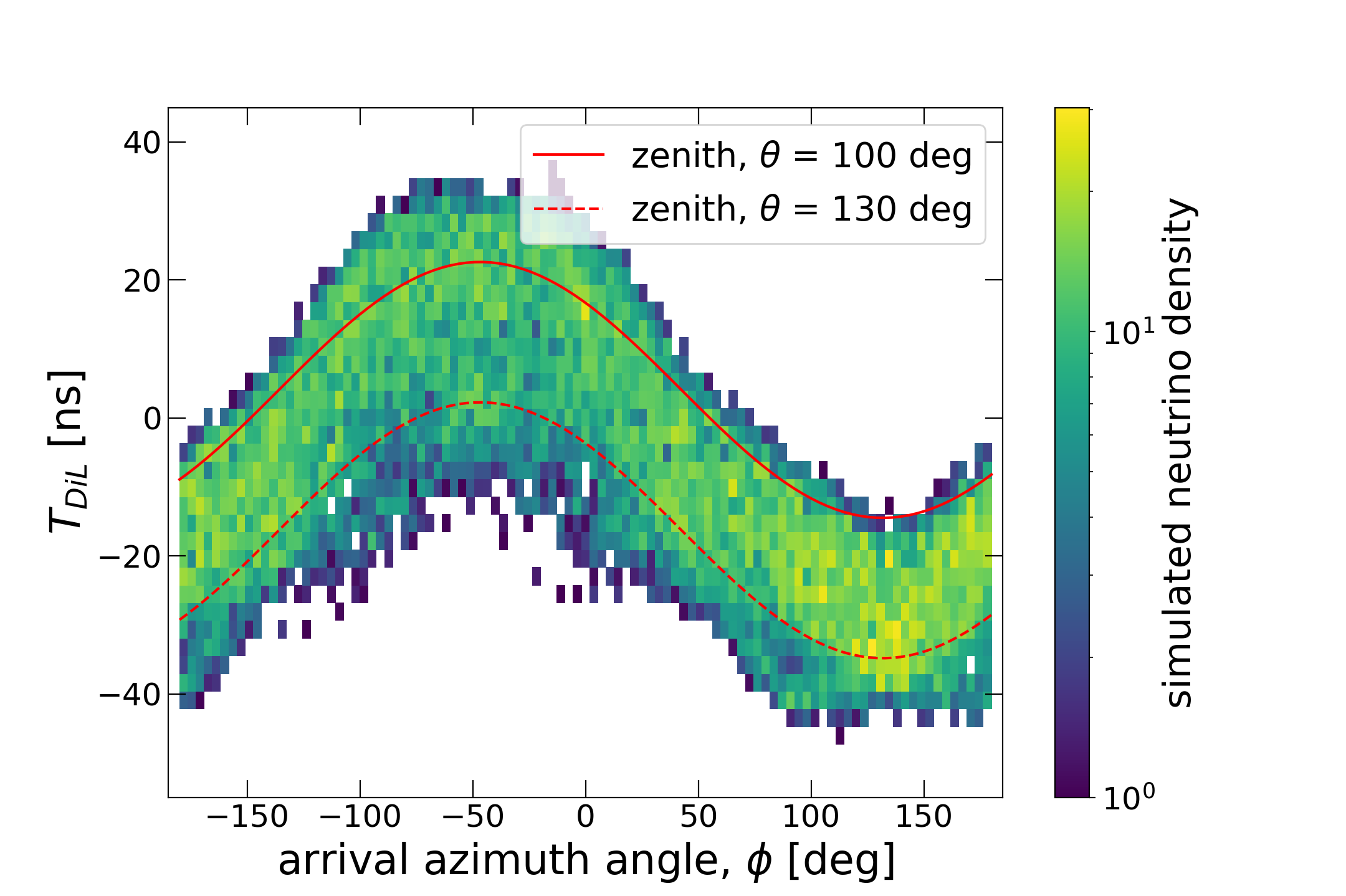}
   \caption{Distribution of $T_{DiL}$, the time interval between the first dipole pulse and the earliest LPDA pulse, for simulated neutrinos as a function of the arrival direction; (left) zenith angle, (right) azimuthal angle.}
  \label{fig:tdlz}
\end{figure}

The set of dipole templates is then cross-correlated with the measured dipole signal, and the template with the highest cross-correlation $\chi_{Di}$ is chosen to be used in further analysis. Then, cross-correlation using the chosen template with the measured signal is performed over a restricted time interval, namely $\pm\SI{10}{ns}$ divided into 100 increments, centered on $T_{DiL}$. Experimental events with large $\chi_{Di}$ values show that the dipole waveform contains (1) the expected number of pulses, (2) $T_{DiL}$ agrees with that obtained from the LPDA reconstruction, \textit{and} (3) the quantities $T_{2P}$ and $V_{pulse1}$/$V_{pulse2}$ in the dipole and template waveforms are comparable. As seen in \autoref{fig:tdlz}, $T_{DiL}$ depends on both the zenith and azimuthal angle of the arrival direction with respect to the ARIANNA station. For a given zenith angle, $T_{DiL}$ varies as a function of azimuth because the horizontal separation varies between the dipole and  nearest LPDA in the signal direction (see \autoref{fig:st_diagrams}).  In addition, the exact time of the maximum amplitude in the LPDA waveform is affected by frequency content of the electric field.  This varies with distance to the interaction vertex and emission angle relative to the Cherenkov cone.  Based on a fit to the simulated neutrino distributions in \autoref{fig:tdlz}, $T_{DiL}$ is calculated from $T_{DiL}=18.5\sin(\phi + 2.39)+27.2\sin(1.5\theta+6.16)-11.89$, where $\phi$ and $\theta$ are azimuthal and zenith angle of the arrival direction in units of radians. Due to the error in this simple fit, the correlation of the dipole waveform, $\chi_{Di}$, is computed over a time interval from $T_{DiL}$-10 ns to $T_{DiL}$+10ns, rather than only at the predicted $T_{DiL}$, and the time that gives the greatest $\chi_{Di}$ is used. It can be seen from \autoref{fig:dt_z} and \autoref{fig:tdlz} that the $\pm \SI{10}{ns}$ flexibility is enough for most neutrinos to account for the deviation of $T_{2P}$ and $T_{DiL}$ from the predicted values due to oversimplification of the template model.

\begin{figure}
\centering
  \includegraphics[scale=0.28]{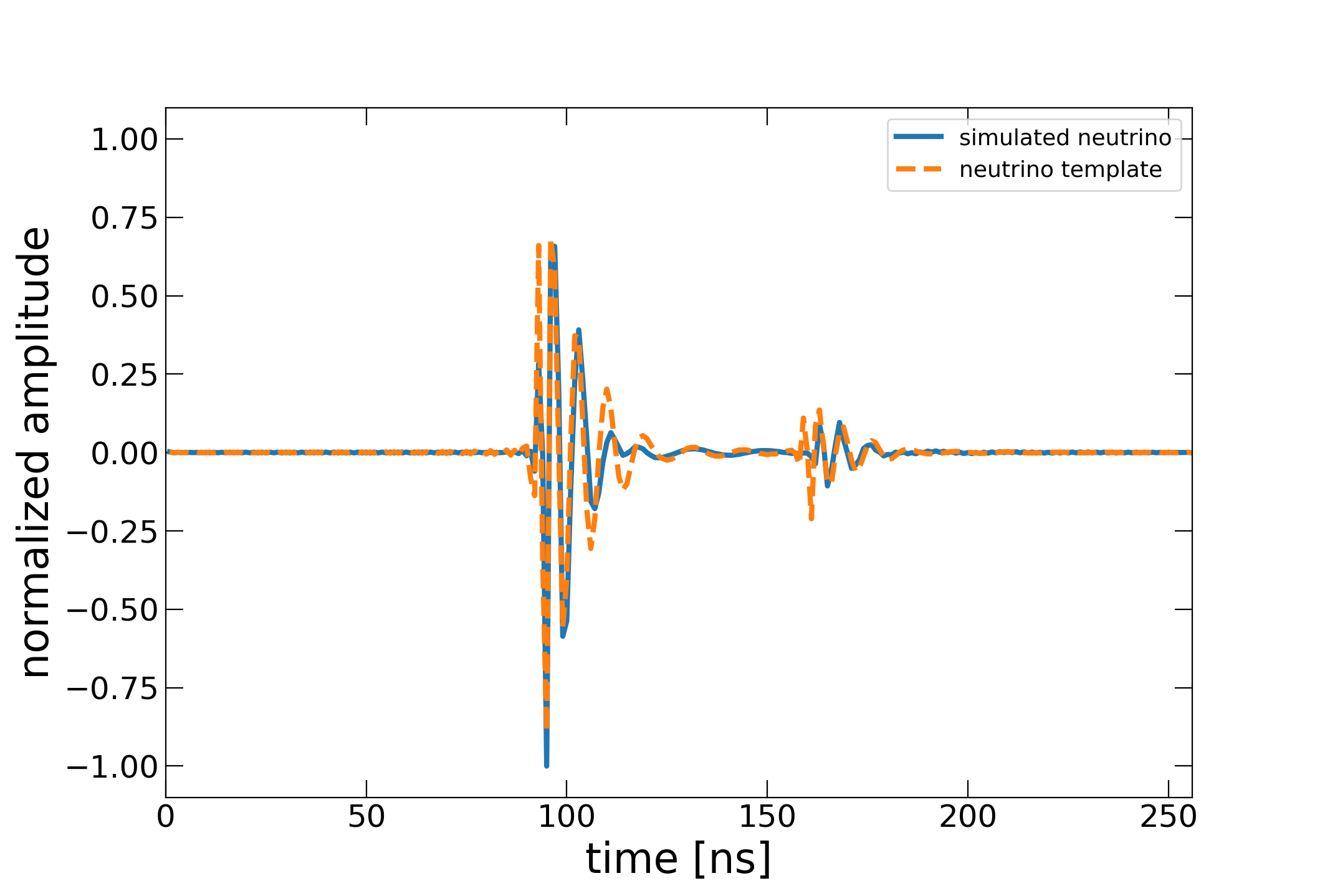}
  \includegraphics[scale=0.28]{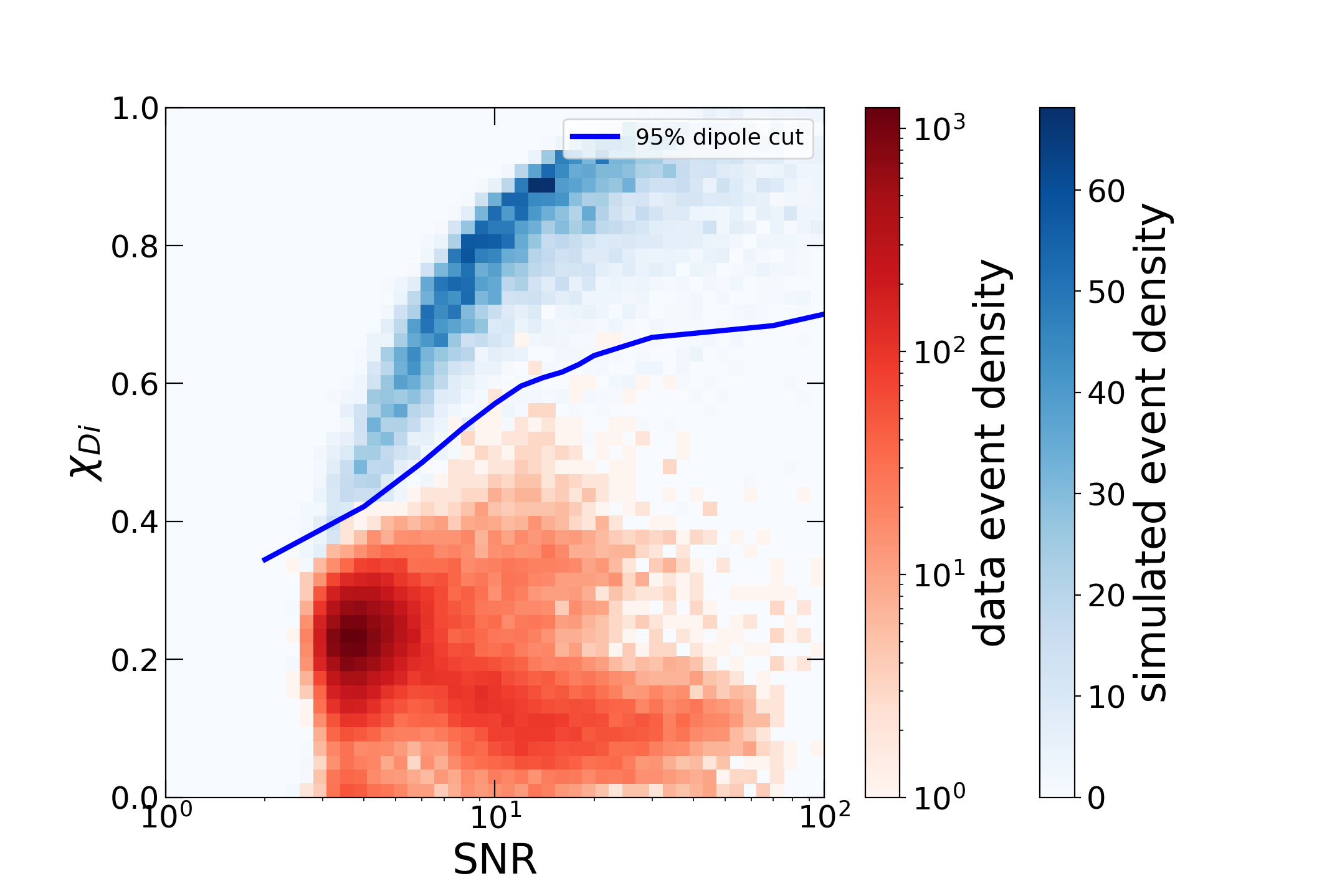}
  \caption{(left:) Comparison between simulated neutrino signal in the dipole antenna (solid blue) and the dipole template (dashed orange) that achieves the best cross-correlation value ($\chi_{Di}$=0.94). The emission angle for the first pulse is \SI{-0.8}{deg} relative to the Cherenkov cone, and its arrival direction is \SI{138}{deg}. (right:) Density plot of $\chi_{Di}$ as a function of signal-to-noise ratio, SNR. The red color legend gives experimental event density  and the blue color legend indicates the simulated neutrino event density. The solid blue line retains about 95\% of the neutrino signal.}
  \label{fig:dipcorr}
\end{figure}

\autoref{fig:dipcorr} shows $\chi_{Di}$ as a function of  the dipole's signal-to-noise ratio (SNR). In this figure, simulated neutrino signals(S-NU61) are shown in blue, and station 61 experimental events (E-BG61) are in red. A greater $\chi_{Di}$ is seen for simulated neutrinos, as expected. Random thermally-induced fluctuations in the waveform  have a greater effect in smaller SNR events, leading to a smaller correlation. There is no such trend seen for experimental events, since the constructed templates in general do not match the shape of experimental events. The dark red oval shaped cluster of events around $\chi_{Di} = 0.2$ is identified as mostly thermal events, since $\chi_{Di} = 0.2$ is a typical value when cross-correlating between two forced trigger events (which contain purely random fluctuations with $V_{rms} \sim$ \SI{10}{mV}).

In addition to the features described above, a strong separation is seen between the simulated neutrinos and the experimental data. The solid blue curve in \autoref{fig:dipcorr} was calculated to retain $\sim$95\% of the weighted neutrino signal. The full experimental data set from station 61 is shown in the red gradient density scale. It is seen that the vast bulk of experimental data is rejected by this cut, leaving only 53 out of 74,530 background events. 

\subsection{LPDA cut}

\begin{figure}[t]
\centering
  \includegraphics[scale=0.6]{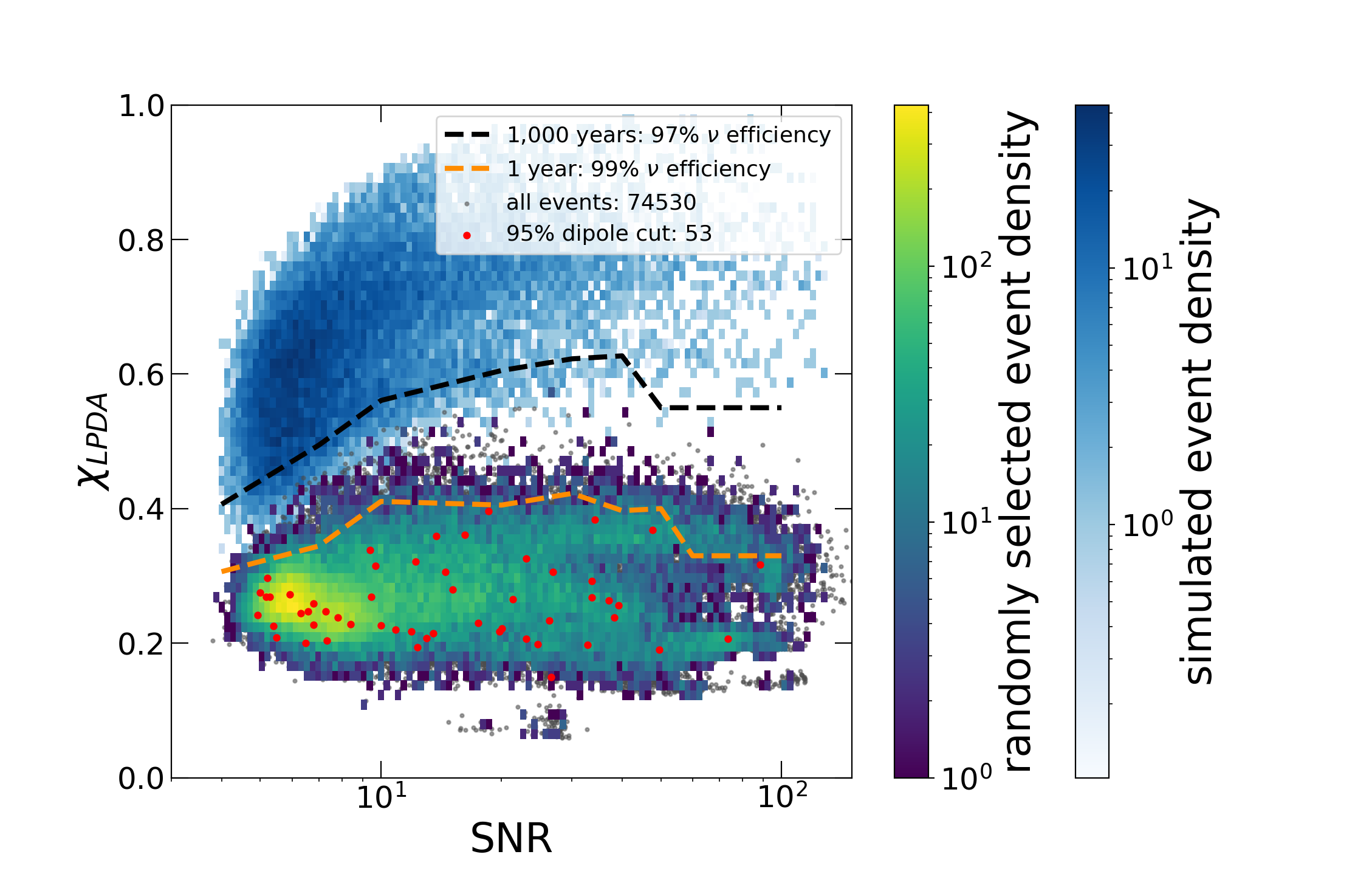}
  \caption{The correlation, $\chi_{\text{LPDA}}$, of observed signals in the  downward facing LPDA in station 61 as a function of signal amplitude, characterized by the largest signal-to-noise ratio (SNR) in the 4 downward facing LPDAs. The neutrino event density is shown in blue, and all experimental data in gray. The color scale shows event densities of experimental data after random selection to scale to 1,000 station-years of livetime. The region above the orange (black) dashed line contains 99\% (97\%) of the neutrino events. The neutrino signal efficiency (in percent) of the individual updown and dipole cuts are given in the legend, along with the combined updown and dipole cut (red dots).}
  \label{fig:lpdacorr}
\end{figure}

For the events that survived the dipole cut, the LPDA correlation cut, $\chi_{\text{LPDA}}$, described in the previously published analysis \cite{Anker:2019rzo} is re-introduced. LPDA waveforms are correlated with a "typical" neutrino LPDA signal template for South Pole ice (in detail, a neutrino signal emitted with an emission angle \SI{1}{deg} from the Cherenkov cone that arrives at the LPDA in a direction \SI{60}{deg} from the nose of the LPDA in the E-plane and \SI{30}{deg} in the H-plane of the antenna). The resulting correlation, $\chi_{\text{LPDA}}$, is shown in \autoref{fig:lpdacorr} as a function of the greatest SNR in any of the four downward pointing LPDAs. Blue data provide the event density of simulated neutrinos, and the color legend indicates the event density of experimental data. The 53 experimental events that pass the updown and dipole cut are shown as red dots. For the 1 year of exposure shown in this plot, the dashed orange curve retains 99\% of the neutrino signal. The dashed black curve is appropriate when extrapolating the exposure to 1,000 station-years (see \autoref{FutureEff}). It retains 97\% of the neutrino signal.

\subsection{Extrapolating the efficiency for a future near-surface radio neutrino detector}
\label{FutureEff}
Based on the data sets and cuts in the preceding section, we calculate the cut efficiency for a future radio-based high energy neutrino array that will run for 1,000 station-years. This live-time consideration is representative of the baseline design for the shallow radio component of IceCube-Gen2 \cite{Gen2Radio2021}. As is pointed out in \autoref{CRid}, the background event populations for ARIANNA stations can be classified as thermal and non-thermal events. Thermal events are generated by random radio emission from the finite temperature ice and/or random fluctuations in the output of the amplifier. Non-thermal events are mostly generated during periods of high wind and by electronic emission from the ARIANNA electronics. In \autoref{fig:lpdacorr}, thermal events dominate in the highest density circular region (yellow), but constitute only about 10\% of the total events. Events that pass the dipole cut are visually consistent with a random distribution within the (predominantly non-thermal) background events. Crucially, they do not cluster on the upper edge of the background distribution closest to the neutrino signal region, which is consistent with the expectation that the dipole, updown, and LPDA cuts are independent. For a background event (assumed to propagate through the atmosphere to the ice surface) to pass the LPDA cut, it should produce a single pulse shape in the backlobe of the horizontally polarized LPDA that has shape consistent with a neutrino signal, whereas background events that survive the dipole cut must produce a double pulse structure in the vertically polarized dipole with the appropriate time delays ($T_{DiL}, T_{2P}$). Assuming that $\chi_{Di}$ and $\chi_{\text{LPDA}}$ for background events are uncorrelated and that the background events observed by station 61 are typical of the future array of shallow detector stations at the South Pole, it is possible to extrapolate the efficiency calculation to an operational livetime of 1,000 station-years. Normalizing the background rate to 53/station-year of operation, which is the rate of events that pass the dipole and updown cut, the estimated background distribution for 1,000 station-years of operation is obtained by randomly selecting a total of 53,000 events from the entire background event population, shown in \autoref{fig:lpdacorr} as the blue-yellow colored population. Since the average amplitude of the observed non-thermal background events is approximately a factor 2 larger than the thresholds in station 61 in the ARIANNA station (see 
\autoref{fig:snr_52}), reducing the thresholds in a future station is unlikely to introduce a new population of non-thermal backgrounds.  For the projected background distribution, the black dashed curve was designed to keep 97\% of the neutrino signal while rejecting all background. 

The combined efficiency of the above three cuts (updown, dipole, and LPDA) for 1,000 station-years is computed from $\epsilon_{tot}=\epsilon_{\text{updown}} \; \epsilon_{\text{dipole}} \; \epsilon_{\text{LPDA}}$ = (0.99)(0.95)(0.97) = 0.91. Future work will optimize the efficiency of the developed cuts, including the deep learning cut introduced in \autoref{sec:nu_search}.
It should also be mentioned that the data of Station 61 represent one of the more challenging data sets for this technique because it includes events from electronic artifacts from the ARIANNA Battery Management Unit.  The cause of these events has been identified and a solution found \cite{mp_thesis}.

\section{Evaluating the neutrino analysis efficiency with Deep Learning}
\label{sec:nu_search}
In contrast to the more traditional techniques used to develop selection criteria in \autoref{sec:eff_cuts}, this section describes the development of a cut using a deep learning approach.

\subsection{Model architecture of the deep learning cut}
\label{sec:mod_struc}
\begin{figure}[t]
\centering
  \includegraphics[scale=0.25]{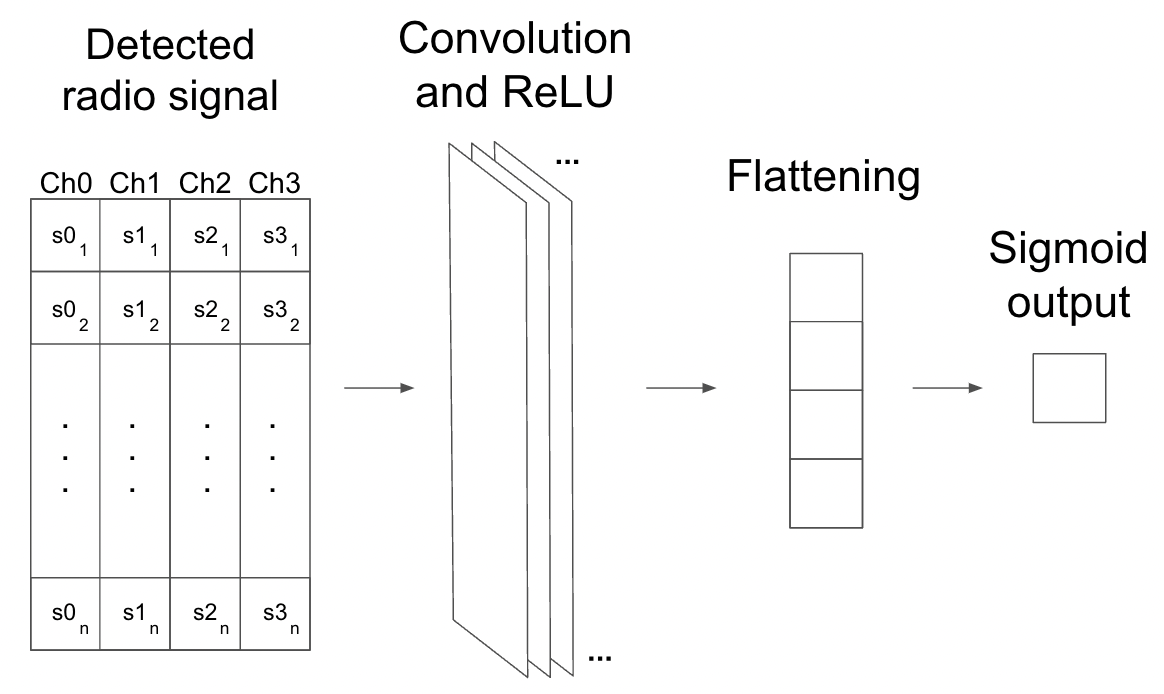}
  \caption{Baseline architecture of a convolutional neural network (CNN). First is a 2D convolution with ReLU activation, then a flatten layer where the data are reshaped, and lastly a sigmoid activation in the output layer.}
  \label{fig:cnn_dgrm}
\end{figure}

The deep learning platform used for training and testing is Keras \cite{chollet2015keras}. The baseline convolutional neural network (CNN) architecture for this study is shown in  \autoref{fig:cnn_dgrm}. Each antenna channel contains 256 samples per event. For station 61 data, all 8 antenna channels were used in the training by stacking them together to produced an input array of 8x256. Comparable, though slightly poorer, results were obtained if only the 4 downward facing LPDA antenna channels were incorporated in the training. However, for the cosmic ray test described in \autoref{CRid}, only the waveforms from the 4 upward facing LPDAs were stacked together, which produced an input array of 4x256, due to concerns about the reliability of the simulated cosmic ray signals in the vertically polarized dipoles and backlobe response of the downward pointing LPDAs. The input data from both stations are run through a convolution with 10 8x10 (8 input channels) or 4x10 (4 input channels) kernels and a ReLU activation. Next the flatten step reshapes the data to a 1D array, and lastly the output layer has a sigmoid activation to give classification values between 0 and 1. This architecture builds on previous work \cite{Anker_2022} in which CNNs were found to be more efficient at discriminating between signal and noise compared to fully connected neural networks. Furthermore, Sherpa \cite{sherpa}, a hyperparameter optimization library for machine learning models, was used to identify the simplest architecture that optimized the classification efficiency by varying the number of epochs, the number of kernels, the kernel size, and the amount of layers; refer to \cite{aanker_thesis} for more details.

\subsection{Using Deep Learning to identify neutrinos}
\begin{figure}[t]
\centering
  \includegraphics[scale=0.5]{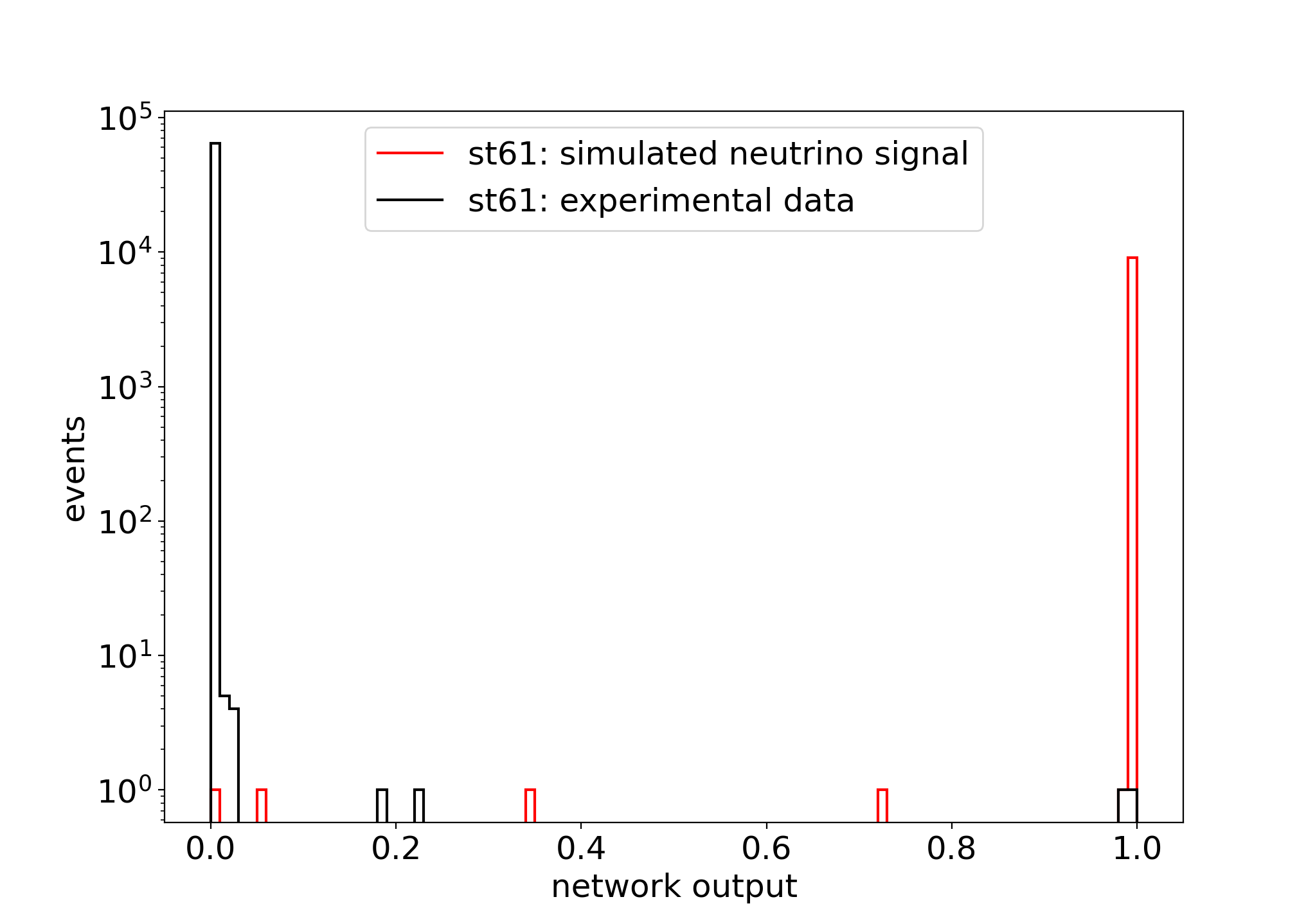}
  \caption{Histogram of the network output for experimental background data and simulated neutrino data for station 61. A network output value close to 0 is experimental data and close to 1 is simulated neutrino signal data.}
  \label{fig:nu_netout}
\end{figure}

 For the deep learning approach, the same station 61 experimental data from \autoref{sec:eff_cuts} is used to evaluate the neutrino efficiency. A network is trained with all of the channels of input data and an architecture consisting of one hidden layer with size 10 8x10 kernels, a flatten layer, and a sigmoid output. There are 10k events from E-BG61, and 1.5k events from S-NU61 for the training data set, and then the CNN is validated with the remaining data. Over multiple trainings of the same model with different initial starting weights, there are two background events that are frequently misclassified; the amount of misclassified events varied between 1 and 4, with these two events appearing most often. These events are representative of the waveforms misclassified by the deep learning cut. The network output of a representative model is shown in \autoref{fig:nu_netout} with the two misclassified events.

\begin{figure}[t]
\centering
  \includegraphics[scale=0.29]{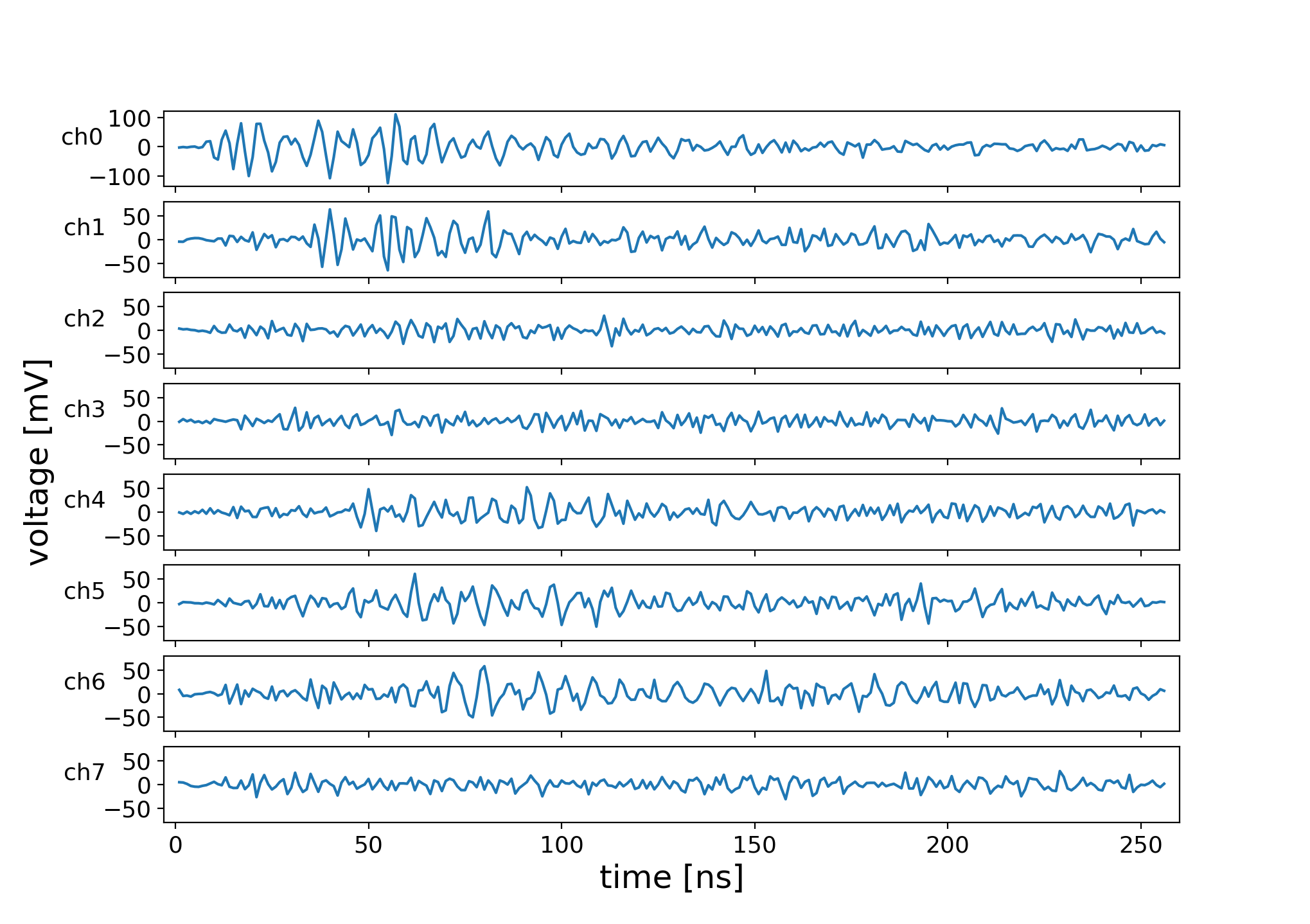}
  \includegraphics[scale=0.29]{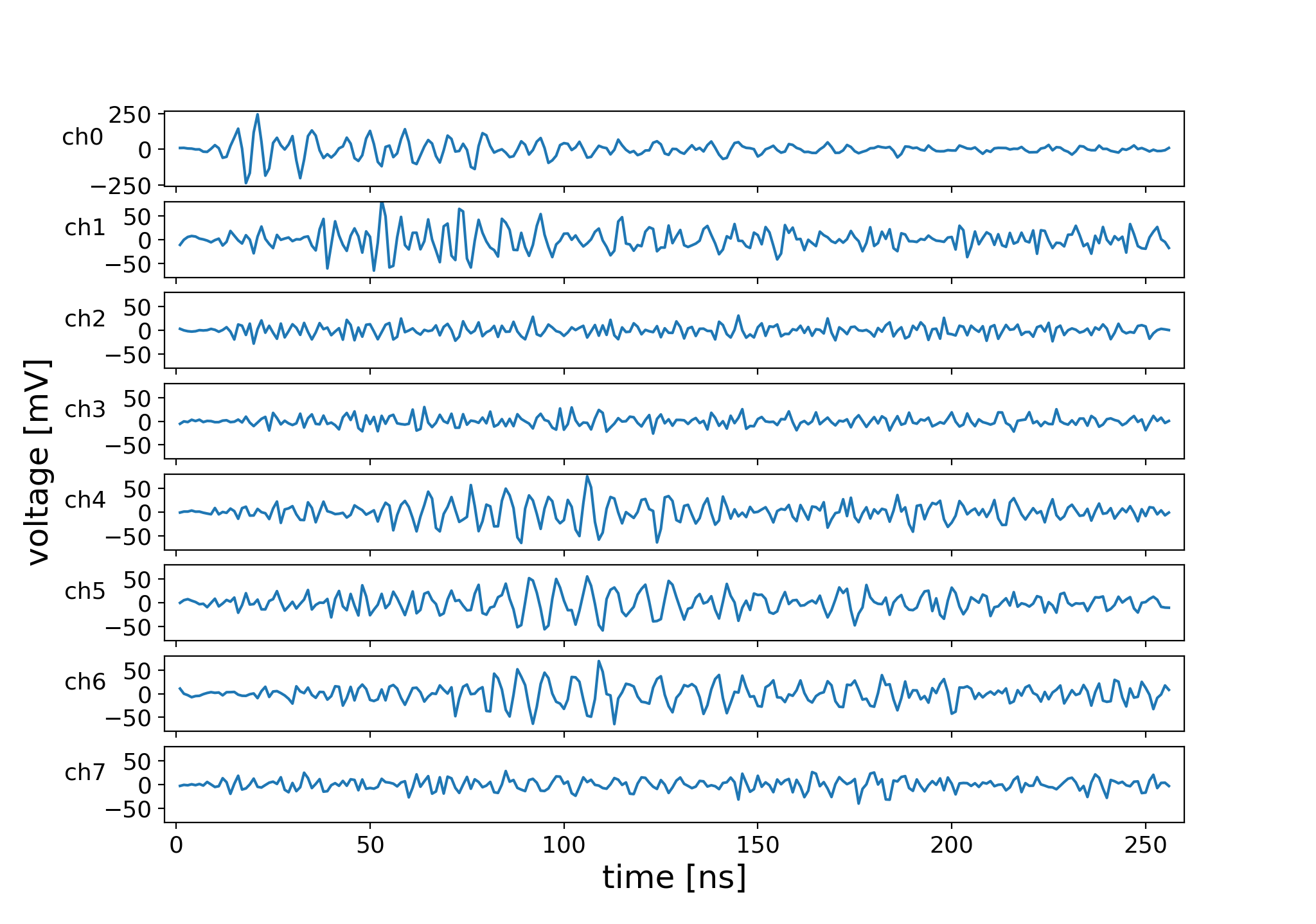}
  \caption{Experimental background events classified incorrectly by the deep learning cut as neutrino signal.}
  \label{fig:two_wfs}
\end{figure}

At a network output cut of 0.5, the CNN achieves almost perfect separation of data with 99.9\% signal efficiency and only two background events in E-BG61 incorrectly identified as S-NU61 events. These two events are shown in \autoref{fig:two_wfs}. A visual inspection of the waveforms in these two events shows that they have strong signals in ch5, connected to an upward facing LPDA, which is not consistent with the expectations of a neutrino event. Also, the channels involved in triggering the detector (0 and 1 in this case) are orthogonal channels whereas in neutrino events, the largest amplitude channels are typically parallel ones. Lastly, the waveforms involved in the trigger extend to $\sim$ \SI{75}{ns}, and for neutrino events the signal pulse is significantly shorter (see \autoref{fig:all_wfs}). Thus, these two events should not be considered neutrino events. Nevertheless, some potential reasons for the misclassification include the uncharacteristically high amplitudes of channel 0 and the more elongated pulse shapes in channels 0 and 1, which look more neutrino-like than most thermal noise events. There are also more high and low pulses in these two events, so it is possible that when the CNN uses kernel correlation to analyze them, it produces more positive contributions to the final network classification.

\begin{figure}[t]
\centering
  \includegraphics[scale=0.6]{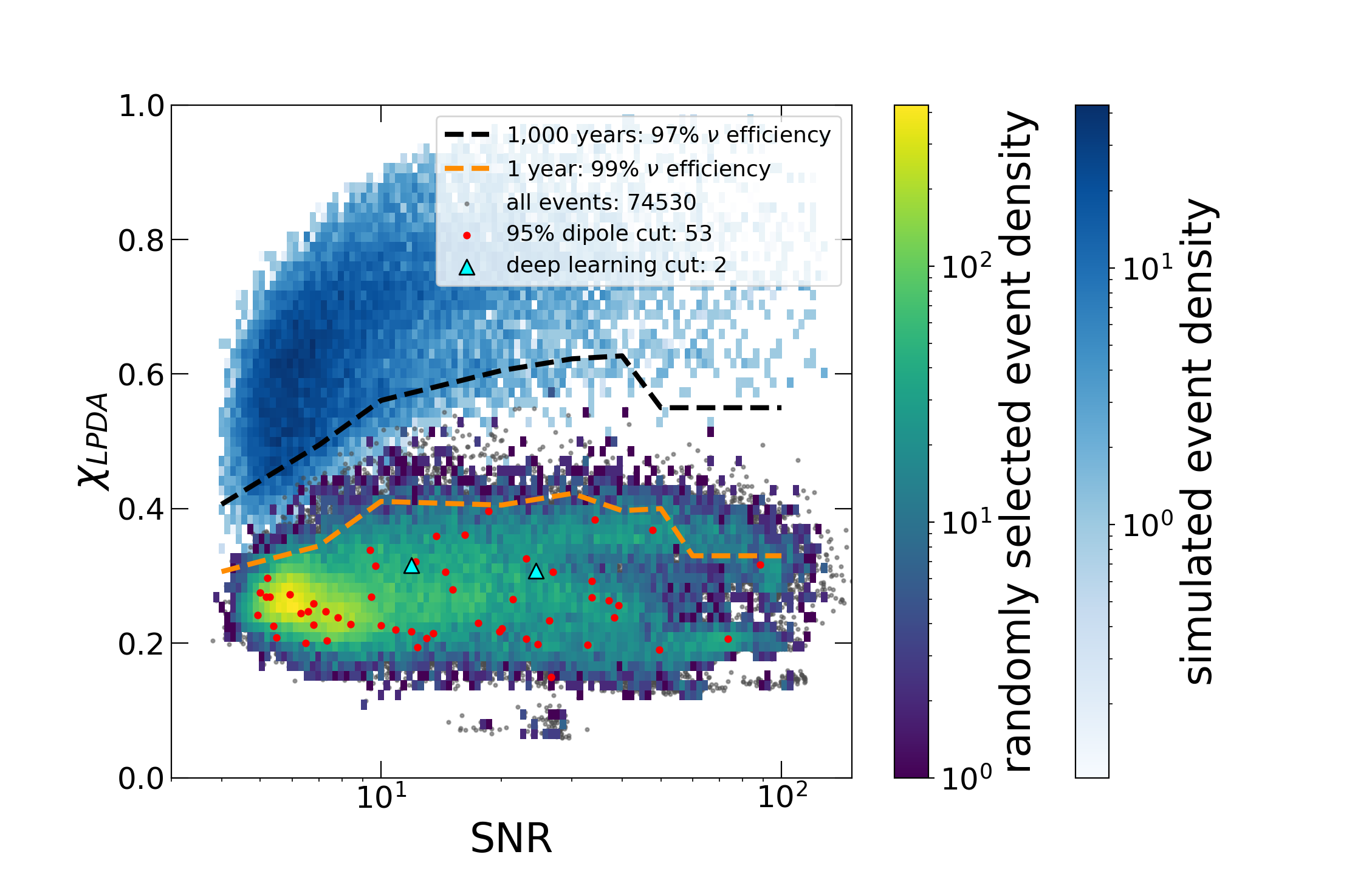}
  \caption{Similar to \autoref{fig:lpdacorr}:  The signal-to-noise ratio (SNR) of the LPDA vs. the correlation of the LPDA experimental data with a simulated LPDA neutrino template. The red dots show the experimental events that pass the updown and dipole cuts. Blue triangles show the two events that pass the deep learning cut .}
  \label{fig:correlate}
\end{figure}

As with the more traditional neutrino search analysis, the LPDA cut is applied to events that pass the deep learning cut.  After this cut, no experimental events remain. The plot of correlation of the LPDAs versus SNR is given in \autoref{fig:correlate}. The events that survive the updown and dipole cuts from \autoref{sec:eff_cuts} in red circles and those that pass the deep learning cut in blue triangles. Also shown are the simulated neutrino signal event density and the experimental noise event density. In particular, the LPDA correlation values of the events that pass the deep learning cut are near the average value for thermal noise events, which is approximately $\chi_{\text{LPDA}}=0.3$. Both the cuts in \autoref{sec:eff_cuts} and the deep learning cut, when combined with the LPDA cut, reject all data events from station 61. Although these two methods give identical results after the correlation cut; the deep learning method rejects over 25 times more events for the same neutrino signal efficiency.

\subsection{Using Deep Learning to identify cosmic rays}
\label{CRid}
\autoref{fig:correlate} indicates the promise of including deep learning techniques in the selection criteria for neutrino identification. However, since the deep learning study in the preceding section includes a mixture of experimental and simulated data, it is possible that the network may have identified an artificial difference between the simulated and experimental data. For example, perhaps a small DC offset in the experimental data is not replicated in the simulated signal events, and this difference could provide an artificial signature for the network to use to distinguish background from signal. Since experimental neutrinos are expected to be rare, and have yet to be measured by the ARIANNA detector, building an experimental data set of neutrino signal is not possible. Fortunately, experimental cosmic rays can be used as a proxy for a training set of experimental neutrino events. Similar to the procedure described in the preceding section, simulated cosmic ray events are used in the training. The experimental data from the cosmic ray station provides the background events for the training set, after cosmic rays, which were identified by an unrelated analysis \cite{lzhao_2022}, are removed. Since artificial differences are not expected between the experimental cosmic rays and the experimental background training set from the same ARIANNA station, a network that depends on an artifact would improperly identify nearly all the experimental cosmic rays as background events. Conversely, if the network properly identifies most experimental cosmic ray events as signal, then this provides evidence that the network is not significantly influenced by artificial differences in the experimental and simulated training sets. 

In contrast to the neutrino study above, the experimental background from the cosmic ray station contains a larger fraction of non-thermal events, for reasons discussed in \autoref{subsection:stn_comp}. Additionally, the frequency content of cosmic ray events is biased toward lower frequencies compared to neutrino events, which is also characteristic of wind and electronic noise events. These two factors caused a large fraction of simulated cosmic rays to be misidentified that is unrelated to the question of possible artificial differences between the simulated cosmic rays and experimental backgrounds. To strengthen this conclusion, the background training set was preprocessed to increase the fraction of thermal noise events. The strong difference in waveform shapes between simulated cosmic rays and experimental thermal events helps to improve the accuracy of network classification. The preprocessing stage is described next.

\begin{figure}[t]
\centering
  \includegraphics[scale=0.5]{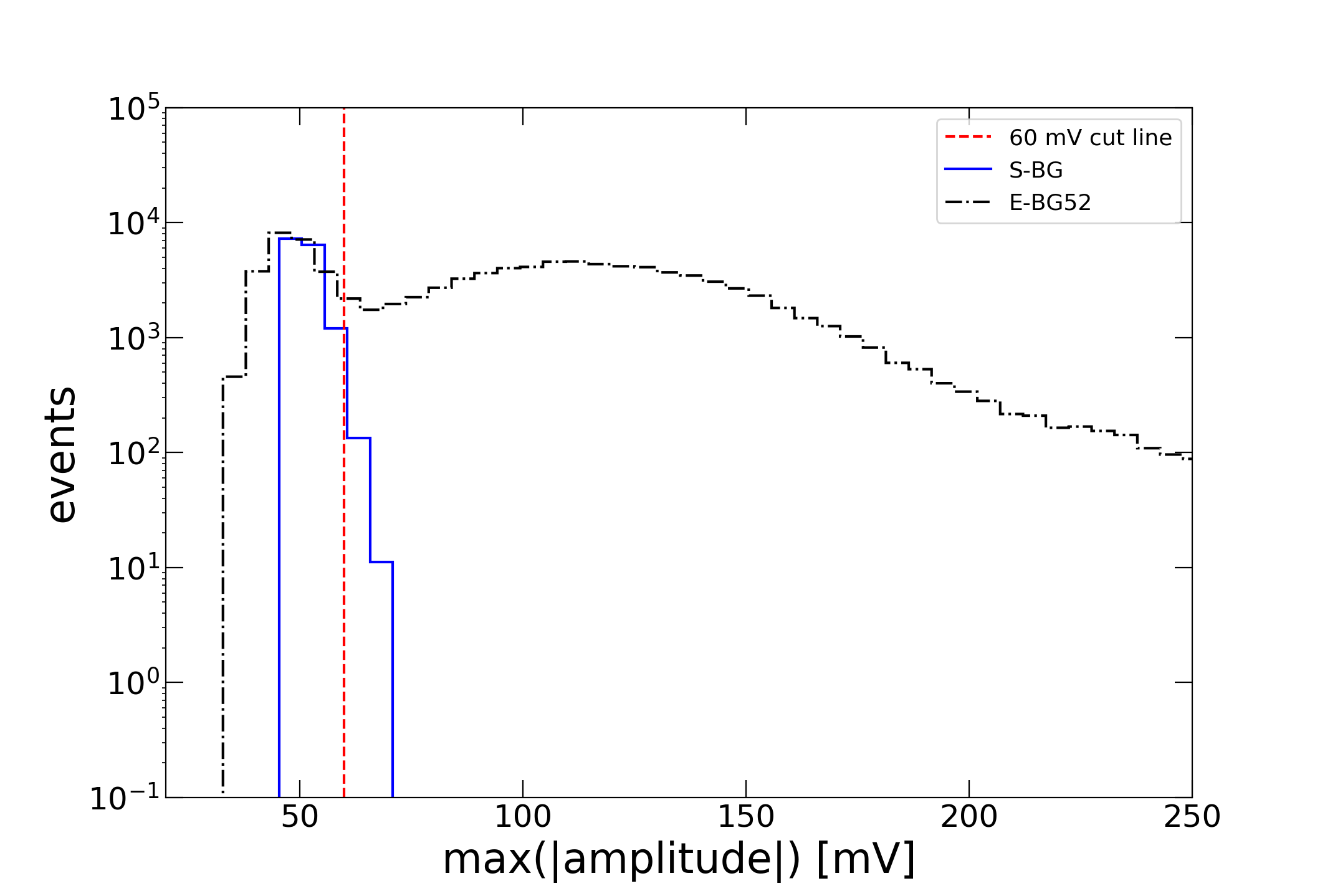}

  \caption{Histogram of the maximum absolute value of the maximum waveform for experimental data collected by the cosmic ray station (E-BG52, black dash dot curve). In addition, simulated thermal events (S-BG, blue solid curve) due to thermally induced fluctuations are shown. Events with amplitudes below the vertical red dashed line at \SI{60}{mV} are dominated by thermal events.}
  \label{fig:snr_52}
\end{figure}

\autoref{fig:snr_52} shows that there are two populations in the experimental data. The low amplitude peak at around \SI{50}{mV} is compatible with the amplitude distribution due to thermal events caused by thermal fluctuations in the signal from the ice and amplifier (blue solid curve). Note that the hard cut of \SI{44}{mV} implemented in the simulated thermal noise events only approximates realistic conditions. During the operation of station 52, the trigger threshold for every downward LPDA channel was slightly different, with an average of \SI{44}{mV}. Additionally, the simulation does not take into account temperature related affects on the threshold value. The second broader peak of non-thermal background events are higher in amplitude compared with thermal noise, generally peaking around \SI{110}{mV}; they are generated during periods of bad weather and high winds. To increase the thermal event fraction in the experimental data, events with max amplitudes below \SI{60}{mV} are used in the training. This procedure retains 23,478 events in the data set E2-BG52.

Next, a mixture of simulated cosmic rays and measured background data was used to train and validate a deep learning model. The data sets for training are E2-BG52 (20,000 out of 23,478 events) and S-CR52 (5,000 out of 9,630 events), which provides adequate statistical precision. The network is trained with 4 channels (upward facing LPDAs) of input data and an architecture consisting of one hidden layer with size 10 4x10 kernels, a flatten layer, and a sigmoid output. The neural network is then validated on the remaining data sets that were not used in training along with the experimental cosmic ray data set of 85 events. 

\begin{figure}[t]
\centering
  \includegraphics[scale=0.5]{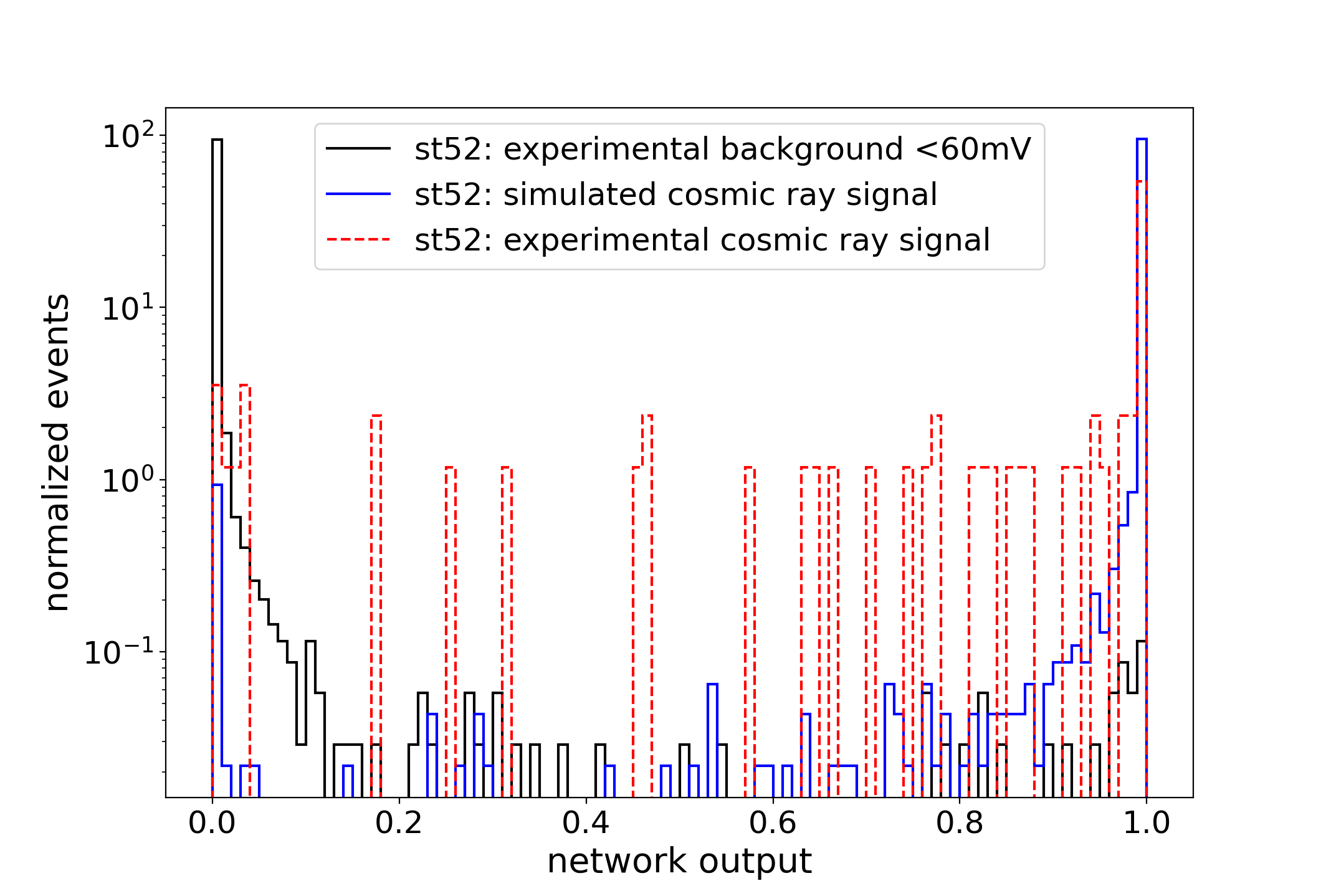}
  \caption{Histogram of the network output for a subset of experimental station 52 data with the maximum amplitudes below \SI{60}{mV}, simulated cosmic rays, and experimental cosmic rays. All distributions are normalized to 85 events to match the experimental cosmic ray data set. The network was trained on the first two data sets mention above. A network output value close to 0 is experimental background and close to 1 is cosmic ray signal.}
  \label{fig:cr_netout}
\end{figure}

The network output histogram distributions are shown in \autoref{fig:cr_netout}. Although there are not many events in the experimental cosmic ray data set, its network output follows a similar distribution to the simulated cosmic ray data set and over 50\% of measured cosmic rays (47 out of 85) are in the bin of the largest network output. If there were irrelevant artifacts in the simulated or experimental data, the network would pick up these differences between simulated and experimental cosmic rays. Since their network output distributions are well correlated, there is no such artifact or feature seen. To further illustrate the similarities between the experimental cosmic ray and simulated cosmic ray distributions, a statistical analysis is performed on the data.

\begin{figure}
\centering
  \includegraphics[scale=0.52]{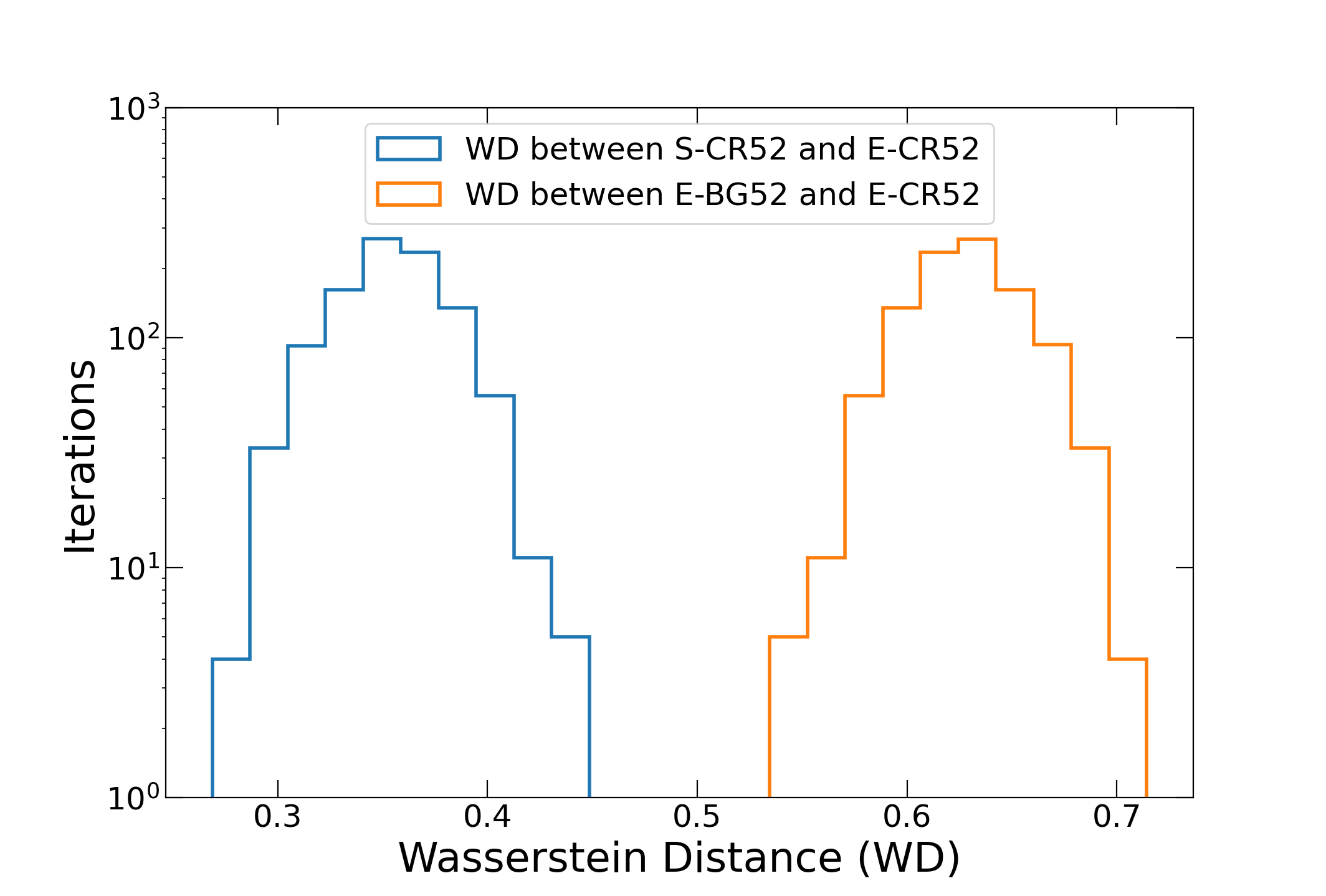}
  \caption{The Wasserstein distance (WD) versus the iterations of data fluctuation (1,000 in total). The blue curve gives the WD between the simulated and experimental cosmic ray distributions. The orange curve gives the WD between the experimental noise and cosmic ray distributions.}
  \label{fig:WD}
\end{figure}

This statistical analysis uses the Wasserstein distance \cite{Villani2009} to compare the experimental cosmic ray distribution to both the simulated cosmic ray and experimental noise distributions. The Wasserstein distance is a mathematical metric for describing the similarity (or distance) between two probability distributions. Only the counts of each bin of the experimental cosmic ray histogram are varied according to a Gaussian function for bins containing greater than four events. For bins containing four or fewer counts, a Poisson distribution is used. The calculation is repeated to produce 1,000 statistically varied histograms. The Wasserstein distance is computed between the simulated cosmic ray distribution and experimental cosmic rays, and also the background noise distribution and experimental cosmic rays for each histogram. The 1,000 computed distances are plotted both for cosmic ray signals and background noise events in \autoref{fig:WD}. The Wasserstein distance is smaller between the two cosmic ray distributions (shown in blue). Thus if the experimental cosmic ray distribution is fluctuated over many iterations, it is unlikely to fluctuate into a distribution that matches the experimental background data.

The cosmic ray study demonstrates that the encouraging results from the deep learning cut for the neutrino analysis can be trusted. It is important to note that the distributions in \autoref{fig:cr_netout} do not show that cosmic rays can be identified efficiently, only that mixing simulated and experimental data when training a model is possible. Further studies and more complex deep learning models would need to be built to identify cosmic rays in either real-time or an offline analysis. However, this is not the main mission in this work; here, the main goal is to measure and identify neutrino signals.

\section{Conclusion and Outlook}
This work described three new analysis cuts that exploited new features in a special-purpose ARIANNA station with 8 antennas that combined a vertically oriented dipole antenna with upward and downward facing LDPAs. These new analysis cuts, when combined with a waveform shape cut on the downward facing LPDA that was described in an earlier paper, remove all background events while retaining 91\% of the neutrino signal if projected to 1,000 station-years of operation. This result presents the first evidence that a detector station with near-surface antennas satisfies the requirements for the baseline design of the radio neutrino component of the IceCube-Gen2 project. The new cuts utilize the unique capabilities of the additional antenna channels. In particular the updown cut relied on the directional asymmetry of the LPDA to differentiate downward traveling radio signals from upward traveling neutrino signals, and the dipole cut relies on the double pulse structure from a neutrino event. These two new cuts, by themselves, reject all but 53 out of 74,530 experimental events from station 61 for one station-year of operation, while retaining 99\% of the neutrino signal. They provide a powerful augmentation to the neutrino identification capabilities of a near surface high energy neutrino detector. Finally, a third powerful new analysis cut was developed using deep learning methods ("deep learning" cut). This cut alone rejected all but 2 out of 74,530 events in the experimental data at 99\% neutrino efficiency per station-year of operation, and no experimental events remained when combined with the waveform shape cut. Thus, it is a factor of 25 more effective at removing backgrounds than the updown and dipole cut combined. We conclude that deep learning methods will play an important role in future radio-based neutrino detectors that anticipate 1,000 station-years of operation or more. Although the neutrino efficiency calculations were based on a widely discussed model of the cosmogenic diffuse flux, these cuts can be applied to steady state and transient point source searches. It is expected that the efficiency will increase due to the less challenging background requirements.  

The training sets for deep learning involved simulated neutrino signals and realistic background events from archival data collected with ARIANNA stations. We then used archival data from an ARIANNA station specifically designed to measure cosmic rays to show that the results from the deep learning analysis were not affected by potential artificial differences between simulated data (from the NuRadioMC code) and experimental data. Note that not all cut combinations were studied in this paper. Future work will assess the statistical independence of the deep learning cut when combined with other cuts, such as the dipole cut, and  develop an optimization procedure to include all of the analysis variables (LPDA cut, updown cut, dipole cut and deep learning cut). Additional future work will include broadening the range of astrophysical spectra and extending the analysis to events with signal to noise as small as 2.

Previous offline analysis of the ARIANNA data \cite{Anker:2019rzo} demonstrated the effectiveness of the LPDA correlation cut in the task of removing background events while retaining a high fraction of neutrino events. That cut relied on the unique \emph{chirped} waveform shape from the LPDA induced by a neutrino event, and contrasted it with the shape of thermal induced triggers that tend to be bipolar pulses with a duration of only a few nanoseconds. The deep learning cut in this paper shows that all but a few background events per station-year can be rejected by a relatively simple CNN. This could have far-reaching consequences. Thus far, simulation studies of future concepts for surface stations have implemented a trigger that diminishes the sensitivity for the dispersed waveforms from an LPDA due to the reduced signal to noise. However, if the unique characteristics of the LPDA waveform can be exploited by analysis tools to increase the efficiency of neutrino detection, as shown in this paper, then the neutrino sensitivity will improve. In addition to the \textit{offline} analysis incorporating deep learning, a CNN was trained to filter out unwanted thermal events from the simple majority logic trigger \cite{Anker_2022}. The events removed by the CNN filter are neither saved nor transmitted over the satellite.  By implementing the CNN filter on the as-built ARIANNA data acquisition electronics, the trigger rate of a surface station with the experimentally proven microprocessor (MBED \cite{mbed1768}), which was chosen for its low power consumption,  increases by a factor $10^3$ (and by a larger factor for more modern microcomputers such as the Raspberry Pi), which directly corresponds to a reduction in threshold. As a practical matter, the CNN filter was undemanding in terms of computer resources and power consumption. Given the initial success of deep learning methods to distinguish neutrino signal from background, and the modest impact on computer resources, it may be transformative to incorporate deep learning methods directly into a real-time trigger \cite{Glaser2023_ICRC} to reduce the trigger threshold and increase the sensitivity.  This idea takes advantage of recent improvements in the digitization speed of low power analog to digital (ADC) devices. Once the digital ADC information is transferred to an FPGA, a real-time trigger based on deep learning networks would efficiently identify and reject most of the unwanted thermal events.

\section{Acknowledgement}
We are grateful to the U.S. National Science Foundation-Office of Polar Programs, the U.S. National Science Foundation-Physics Division (grant NSF-1607719) for supporting the ARIANNA array at Moore's Bay,and NSF grant NRT 1633631. This work was also supported by the U.S. Department of Energy under Contract No. DE-AC02-05CH11231. Without the invaluable contributions of the people at McMurdo Station, the ARIANNA stations would have never been built. We acknowledge funding from the German research foundation (DFG) under grants GL 914/1-1 and NE 2031/2-1, the Taiwan Ministry of Science and Technology, and the Swedish Government strategic program Stand Up for Energy. The computations and data handling were supported by resources provided by the Swedish National Infrastructure for Computing (SNIC) at UPPMAX partially funded by the Swedish Research Council through grant agreement no. 2018-05973.

\bibliographystyle{JHEP}
\bibliography{bib}

\end{document}